# Novel Theory of the Structure of Elementary Particles

H. Rose

Ulm University, Albert-Einstein-Allee 11, 89081 Ulm, Germany

Email: harald.rose@uni-ulm.de

## Abstract

A novel theory of the structure of elementary particles is outlined which differs from the Standard Model and the Dirac theory of the electron. The proposed relativistic covariant space-time approach supposes that all massive particles are composite particles formed by massless elementary particles with opposite four-dimensional (4D) helicity. The attraction between two such particles originates from their mutual 4D density, which depends only on their 4D distance. Although this idea take very few seriously, the approach enables a consistent description of the internal structure of massive elementary particles including the origin of their spin, their mass, and the sign of their charge. The new model treats massive elementary particles as four-dimensional (4D) dynamic tops. The associated 4D rotational Hamiltonian depends on the hyper-symmetric potential obtained by using the mutual density for the source term of the 4D Poisson equation. The resulting eigenfunctions of the rotational Hamiltonian depend on the 4D radius $R$ and on three angles, one of which is imaginary. The imaginary angle accounts for the rotation of the time-like axis with respect to the three-dimensional subspace representing a Lorentz transformation. We obtain the wave equation of the basic particles by supposing that they are propagating with the speed of light. Such subatomic massless particles are photons, basic-quarks ("quarkinos"), and neutrinos, which we assume to be massless if they are free. The results of the analytical calculations show that massive particles can only be stable if the potential is a 4D well trapping the constituents. As a result, we do not need the Higgs field for explaining the origin of the mass. Moreover, this potential may also account for quantum entanglement of two photons separated by arbitrary 3D distances. In particular, we show that electrons and positrons are such two-component systems each consisting of two photons. The eigenvalue of the 4D rotational energy determines the mass of the composite particle or its 4D radius, respectively. We prove the relevance of the novel theory by revisiting the hydrogen atom. The resulting formula for the energy eigenvalues depends on three quantum numbers having no degenerated energy states. In particular, it accounts for the Lamb shift, which is unknown to the Dirac theory. Moreover, our approach suggests that the proton may consist of massless spin-1/2 quarks with opposite unit charge and that excited 4D states represent massive quarks.

## 1. Introduction

Albert Einstein stated that the most inconceivable property of nature is the simplicity of its basic physical laws. For this reason he rejected the probability interpretation of quantum mechanics because he was convinced that God "does not toss the dice" [1]. Most problems in understanding complicated physical phenomena result from the fact that we do not exactly know the fundamental underlying physical principles. If we could explain the nature of elementary particles, we would also know the evolution of the universe because it follows the same fundamental physical principles. Each elementary particle behaves as a particle or as a wave depending on the kind of observation. We suggest that all massive elementary particles including electrons and positrons have a substructure and that they are composed of elementary massless constituents, such as photons, neutrinos, and basic quarks (*quarkinos*), which are massless quarks if they exist. We shall investigate this possibility in detail by employing a Lorentz-invariant treatment, which splits the 4D Lorentz-invariant Hamiltonian of a massive particle in a 4D translational term, a 4D rotational term, and an interaction term. This term accounts for the interaction between the intrinsic massless sub-particles or the interaction of the compound particle with an external electromagnetic field. The calculations show that the eigenvalue of the 4D rotational Hamilton operator represents the 4D moment of inertia, which is proportional to the square of the mass of the particle. This result is rather revolutionary because our treatment does not require the Higgs field for explaining the mass of a particle, the electron in particular. The Standard model supposes that the physical masses arise when the gauge symmetry is spontaneously broken resulting in two pieces, one of which is the Yukawa coupling to the Higgs field, the other a pure Fermion mass. Our approach does not require such a splitting. The claim, that the Higgs particle is the source of all masses because its field is imparting an effective inertia to everything that moves, is largely heuristic and to our opinion not proven by the result of the CERN experiments. Moreover, a spontaneous splitting (action without cause) is hard to conceive.

Relativistic quantum field theory and quantum electrodynamics suffer from inherent difficulties due to divergences arising for example by computing the electron self-energy. The divergence difficulties are of such a profound character that they necessitate a drastic change or extension of the present theory. Therefore, we propose that the description of the structure of elementary particles and atomic nuclei requires novel methods, which differ from those employed presently. The new procedures, which we propose, explain the existence of two kinds of particles, bosons with integer spin and fermions with half-integer spin and give detailed information on the four-dimensional structure of elementary particles. The spin of the particle does not represent the 4D angular momentum but only its space-like component. Contrary to the Dirac theory, we postulate that particles exhibiting an intrinsic angular momentum cannot be point-like but must have a finite four-dimensional radius. This requirement is in accordance with the laws of classical mechanics according to which the rotational energy diverges at the limit $r \to 0$. Nature avoids this divergence by means of a "calm eye" realized in hurricanes, tornados, and eddies riding on water waves. The smaller the eye, the stronger is the rotation or the wind, respectively. Moreover, the Dirac theory faces the problem that there is no lower bound to the negative energy spectrum. The heuristic concept of the "Dirac Sea" avoids the catastrophic transitions to infinitely large negative energies by postulating that all negative energy states are filled.

In the following, we shall show that the extension of the space from three to four dimensions offers a promising method for tackling many unsolved fundamental problems and questions. The four-dimensional Minkowski space consists of two sub-spaces separated from each other by the light cone. Two of the subspaces have time-like properties the other two have space-like properties. The part of the universe to which we have access is located within the time-like subspace. The metric coefficients $g_\mu, \mu = 1,2,3,4$, of the "pseudo-Euclidian" Minkowski space are

$$g_1 = g_2 = g_3 = 1, \ g_4 = i. \tag{1}$$

We conceive the imaginary time-like coordinate $x_4 = g_4 ct = ict$ as the fourth spatial coordinate of the "pseudo-Euclidian" Minkowski space, which is composed of the real coordinates $x_1 = x, x_2 = y, x_3 = z$ of the three-dimensional subspace and the imaginary "time-like" spatial coordinate $x_4 = ict$. The coordinates $x_\mu$ can adopt positive and negative values.

The 4D Minkowski space is composed of four three-dimensional sub-spaces, which are orthogonal to each other. The direction of each normal defines an axis in the 4D space. Hence, four orthogonal axes do exist, each of which connects two adjacent sub-subspaces. Twelve surfaces enclose the four-dimensional cube, whereas six surfaces enclose the three-dimensional cube. Owing to this difference, we need an angle of $4\pi$ to achieve a complete revolution within any hyperplane plane in the 4D-space. Moreover, we cannot define rotations in the 4D-space by an axial vector but must describe them by an antisymmetric tensor composed of six components $T_{\mu\nu} = -T_{\nu\mu}$, three of which are real and three imaginary, as it is the case for the components $F_{\mu\nu}$ of the electromagnetic field tensor. The Lorentz-invariant universal time $\tau$ governs the dynamics of particles moving in the time-like region of the four-dimensional pseudo-Euclidian space. Moreover, we assume that the universal time increases monotonically. This assumption is an extension of Newton's conception from the three-dimensional to the four-dimensional space. The universal time should not confused with the "proper time" time of a particle because the universal time is an independent Lorentz-invariant variable, which is the same for all particles regardless of their motion.

The extension of the space from three to four spatial dimensions leads to a change of the properties of physical quantities. We show the result of this transition in table 1 for different physical quantities. By reducing the space from four to three dimensions, we use the laboratory time $t = x_4/ic$ as the independent variable on the expense that this time changes under Lorentz transformations. Accordingly, neither an absolute "now" nor an absolute geometrical space exists. Moreover, there seems to be no universal present moment, because one measures the laboratory time *t* by a clock in the rest frame of the observer. Since we can only measure the time in this system, we cannot determine a universal present moment.

Therefore, the question remains if a universal time $\tau$ exists. Because this time must be a Lorentz scalar, we can speculate that its conjugate momentum variable relates to the mass. Hence, without mass there will be no universal time. Therefore, the universal time must start with the origin of mass formed at the Big Bang and be the same for all particles created at this time. Our definition differs somewhat from that of the proper time, which relates to a single particle. Moreover, we can conceive the universal time as a hidden Bell parameter with a realistic physical property [2]. As pointed out by D. Kwiat [3], the violation of Bell's inequality does not rule out the validity of hidden variables. Entanglement results from the simultaneous creation of two particles, which will keep their original state in the absence of external interference.

In the past, one considered the universal time (proper time) merely as a convenient Lorentz-invariant calculus parameter. Contrary to this view, we assume the existence of a true universal time that governs the dynamics of massive particles in the four-dimensional Minkowski space [4]. As consequence of our postulates, massless particles such as photons and neutrinos are bound to four-dimensional planes, which do not propagate in the pseudo-Euclidian Minkowski space.

| 3D space | | 4D (pseudo-Euklidian) space |
|---|---|---|
| 3 coordinates: $x_\nu, \nu = 1, 2, 3$ | → | 4 coord.: $x_\nu, \nu = 1, 2, 3$, $x_4 = ict$ |
| Polar vector: $\vec{r} = \sum_{\nu=1}^{3} x_\nu \vec{e}_\nu$ | → | Polar vector: $\vec{R} = \sum_{\nu=1}^{4} x_\nu \vec{e}_\nu$ |
| Square: $\vec{r}^2 = r^2 > 0$ | → | $\vec{R}^2 = R^2 \to \begin{cases} > 0 & \text{space-like, closed sytem} \\ = 0 & \text{light-like, radiation} \\ < 0 & \text{time-like, open system} \end{cases}$ |
| Axial vector: $(\vec{a} \times \vec{b})$ | → | Antisymmetric tensor $\overleftrightarrow{T}$ $T_{\mu\nu} = -T_{\nu\mu}$, $\mu, \nu = 1,2,3,4$ |
| Laboratory time: $t$ | → | Lorentz-invariant universal time $\tau$ |
| Hamilton operator: $H_3(\frac{2m}{\hbar^2})$ | → | Mass-independent 4D operator: $\widetilde{H}_4$ |

Table 1: Transition of three-dimensional quantities by extending the space from three up to four dimensions

We classify the properties of four-vectors and antisymmetric second-rank four-tensors $\overleftrightarrow{T}$ according to the sign of their square. Examples of four vectors are the 4D radius vector $\vec{R}$ and the 4D kinetic-moment vector $\vec{P}$, whose square we define as the 4D kinetic energy $E_k^{(4)}$. By writing $\vec{R}$ and $\vec{P}$ as sums of their components, their squares are

$$\vec{R}^2 = \sum_{\nu=1}^{4} x_\nu^2 = \vec{r}^2 - c^2 t^2 = R^2, \qquad \vec{P}^2 = m^2 \sum_{\nu=1}^{4} \left(\frac{dx_\nu}{d\tau}\right)^2 = \vec{p}^2 - p_4^2 = E_k^{(4)}. \qquad (2)$$

In the absence of external forces, the square $P^2 = E_k^{(4)} = -m^2 c^2$ of the 4D kinetic momentum of the particle is a negative constant where *m* is the rest mass of the particle. In the case $R^2 < 0$ and $E_k^{(4)} < 0$ the 4D vectors $\vec{R}$ and $\vec{P}$ are *time-like* indicating that the particle is propagating in an open system. If the square of the 4D vector is positive, its property is *space-like* defining a closed system. The time-like region of the Minkowski space has a hyperbolic metric ($R^2 < 0$), which implies an open universe. Positive values of $R^2 > 0$ are located within the space-like regime of the Minkowski space and are attributed to closed systems, as it is the case for massive particles and eventually black holes. Usually, one excludes the space-like region because $R^2 > 0$ violates classical causality. However, this argument does not hold in quantum mechanics, which allows the wave to penetrate into the classically forbidden region within distances in the order of the wavelength. Accordingly, quantum mechanics allows the existence of closed systems located within the space-like region $0 < R^2 \leq R_0^2$ near the light cone $R^2 = 0$. The light cone separates the time-like regime of the 4D-space from the space-like region. We shall show that the elementary radius $R_0$ relates to the Compton wavelength of the particle.

Most physically relevant four-tensors are antisymmetric. Three of their six independent components $T_{\mu\nu}$ are real and have space-like properties. The other three are imaginary and have time-like properties. By writing the antisymmetric tensor in terms of its components, we obtain for its square the relation

$$\overleftrightarrow{T}^2 = \sum_{\mu=2}^{4} \sum_{\nu=1}^{4} T_{\mu\nu}^2 = 2 \sum_{\mu=2}^{4} \sum_{\nu=1}^{\mu-1} T_{\mu\nu}^2 = T^2. \qquad (3)$$

Electromagnetic waves have light-like property because their electromagnetic field tensor satisfies the relation $\overleftrightarrow{F}^2 = \sum_{\mu=2}^{4} \sum_{\nu=1}^{\mu-1} F_{\mu\nu}^2 = \vec{B}^2 - \vec{E}^2/c^2 = 0$.

Within the frame of our model, we assume that elementary particles are four-dimensional rotating tops moving in Minkowski space. Together with the assumption that massive particles have a finite four-dimensional radius, we avoid divergences and obtain a physical understanding of the internal structure of the elementary particles. Our model supposes that the basic particles are massless and that particles with opposite four-dim3nsional angular momentum articles attract each other by four-dimensional potentials forming massive particles. We restrict our considerations to massive particles composed of two mass-less particles. Massless particles are photons, free neutrinos if they propagate with the velocity of light, and pseudo-quarks. We shall discuss the problem of neutrino masses and the existence of spin-less pseudo quarks in detail. Our procedure enables an analytical treatment of particles consisting of elementary particles and their antiparticles. The results provide a rather precise model for the four-dimensional structure of the electron and the proton. Within the frame of our approach, the electron and positron consist each of two photons with opposite helicity, whereas the proton may consist of two massless quark-antiquark pairs or of two massive quarks, respectively. Although the results of our approach are in conflict with the postulations of the Standard model, it offers a promising alternative because it does not require fractional charges and elucidates the origin of mass, spin, and the sign of the charge.

2. Relativistic wave equation

Photons and neutrinos are mass-less particles which have an intrinsic angular momentum or spin, respectively. The absolute value of the spin is either an integer or a half-integer of the Planck constant $\hbar$. The Standard Model of elementary particles assumes that the starting masses of the quarks and the leptons are zero and that they obtain their mass by coupling with the field of the Higgs particle [5, 6, 7]. We propose an alternative model, which postulates that stable massive elementary are four-dimensional spinning tops, which are formed by the coupling of mass-less particles with their antiparticles. Particle and antiparticle differ from each other by their opposite four-dimensional angular momentum or 4D helicity, respectively. Because they are independent particles, they do not annihilate each other, as it is the case for photons and neutrinos. Our model describes the evolution of the relativistic wave function $\psi$ in the four-dimensional space with respect to the universal time $\tau$ by extending the Schrödinger equation from three to four-spatial dimensions, where the time-like coordinate is imaginary. The dynamics in the four-dimensional space is then a function of the Lorentz-invariant universal time $\tau$, which we do not conceive as the "eigentime" of the particle. Moreover, we postulate that the operator $\partial/\partial\tau$ accounts for the creation or annihilation of particles, which can be massive or massless, such as photons and neutrinos. By supposing that this interpretation is realistic, the quantization of the relativistic covariant 4D Hamilton-Jacobi equation [4] gives

$$\frac{1}{D_0}\frac{\partial \psi}{\partial \tau} - \widetilde{H}_4 \psi = 0, \qquad \widetilde{H}_4 = H_4/\hbar^2. \tag{4}$$

Here $\widetilde{H}_4$ is the modified Lorentz-invariant 4D Hamilton operator. Equation (4) differs from the non-relativistic Schrödinger equation in so far that the imaginary unit $i$ is missing on the left side and that we have substituted the covariant 4D Hamilton operator for the 3D Hamilton operator. The reason for omitting $i$ is that $\tau$ is not a time-like coordinate of the 4D space but an independent Lorentz-invariant variable, which accounts for the evolution of the quantum system in the 4D space regardless of the specific inertial system. Moreover, the four-dimensional Hamilton operator needs not to be a real function as it is the case for the Schrödinger equation.

Equation (4) has the structure of a four-dimensional diffusion equation with an elementary diffusion constant $D_0 = c/k_0$. By supposing that this relation is realistic, the diffusion constant defines a yet unknown elementary wavelength $\lambda_0 = 2\pi/k_0$ or elementary time $\tau_0 = \lambda_0/c$, respectively. We can conceive the time quantum as the duration of the presence. Equation (4) depends linearly on the derivative $\partial \psi / \partial \tau$. Therefore, the equation differentiates between past and future with respect to the universal time $\tau$ but not with respect to the laboratory time *t*, as it is the case for the Schrödinger equation and the Dirac equation. Accordingly, only events at previous universal times $\tau' \leq \tau$ have an effect at time $\tau$. This reasoning does not hold for the laboratory time *t*. In case that the 4D Hamilton operator $H_4$ does not depend on the universal time, the state is either static ($\partial \psi / \partial \tau = 0$) or stationary ($\partial \psi / \partial \tau = i\omega \psi,$). In the static case, $H_4$ is real whereas it is complex (non-Hermitian) in the stationary case. The imaginary part results either from continuous radiative interactions with external fields and/or is due to the existence of a Lorentz-invariant frequency quantum $\omega_0$ of oscillating particles. By extending the space from three to four dimensions, we avoid the need to linearize the Klein-Gordon equation, as performed by Dirac. Instead we extent the Schrödinger equation from three to four spatial dimensions and substitute the universal time $\tau$ for the laboratory time *t* in the linear time-operator. This procedure has the advantage that it is not restricted to a single particle, as it is the case for the Dirac linearization procedure. Moreover, our approach employs a 4D Hamiltonian, which takes into account the intrinsic structure of non-point-like elementary particles.

The modified Hamiltonian operator $\widetilde{H}_4$ does not depend explicitly on the mass of the particles. Therefore, it is valid for massive particles and for massless particles such as photons and neutrinos provided that they are massless. The particle propagates within the time-like region of the Minkowski space if it is massive and on the light cone if it is massless. Instead of breaking the symmetry, we postulate that the total energy of the particle is composed of its four-dimensional translational energy, its 4D rotational energy, and the energy resulting from its interaction with an external field. Our relativistic covariant approach guarantees that the modified 4D Hamilton operator is Lorentz-invariant having the form

$$\widetilde{H}_4 = \widetilde{H}_t + \widetilde{H}_r + \widetilde{H}_i . \qquad (5)$$

We obtain the terms of the operator (5) by extending the classical quantization procedure from three to four dimensions. The term $\widetilde{H}_t$ accounts for the translational motion, the term $\widetilde{H}_r$ for the rotation of the particles, and the term $\widetilde{H}_i$ for the interactions between the between particles and/or for the interaction with external fields.

In the case of a single massless particle propagating in free space, the interaction operator $\widetilde{H}_i$ nullifies, and the translational operator $\widetilde{H}_t$ and the rotational operator $\widetilde{H}_r$ adopt the form

$$\widetilde{H}_t = \frac{1}{\hbar^2}\sum_{\mu=1}^{4} p_{k\mu}^2 = -\sum_{\mu=1}^{4}\frac{\partial^2}{\partial x_\mu^2}, \qquad \widetilde{H}_r = -\Delta_r^{(4)} = T_R + \frac{\overleftrightarrow{M^2}}{\hbar^2 R^2}, \qquad (6)$$

$$T_R = -\frac{\partial^2}{\partial R^2} - \frac{3}{R}\frac{\partial}{\partial R}, \qquad \overleftrightarrow{M^2} = \sum_{\mu,\nu} M_{\mu\nu}^2 = -\hbar^2 \overleftrightarrow{\Omega^2}, \qquad (7)$$

$$M_{\mu\nu} = p_\mu x_\nu - p_\nu x_\mu = i\hbar\left(x_\mu \frac{\partial}{\partial x_\nu} - x_\mu \frac{\partial}{\partial x_\nu}\right) = i\hbar\Omega_{\mu\nu}(\alpha,\theta,\phi). \qquad (8)$$

The Dirac theory assumes a point-like electron and accounts for its 4D rotation by a spinor wave function consisting of four components $\psi_\mu(x_\nu)$. Contrary to the Dirac theory, our approach describes the particle by a single function $\psi = \psi(x_\nu, R, \alpha, \theta, \phi; \tau)$ depending on nine variables. The four coordinates $x_\nu, \nu = 1,2,3,4$, define the position of the center of gravity of the particle and $R, \alpha, \theta, \phi$ the position of a point with respect to the center $R = 0$ of the non-point-like particle. The Lorentz invariant tensor operator $\overleftrightarrow{M}$ is the 4D equivalent of the three-dimensional angular-momentum Hamilton operator. The operator $T_R$ accounts for the kinetic energy of the motion along the 4D radius. As a result, the particle can expand or shrink its size by interaction with external fields. The four-dimensional angular-momentum operator $\overleftrightarrow{M}$ is an anti-symmetric second-rank tensor operator, which has six linearly independent components. We determine these components by introducing hyper-spherical coordinates.

In the case of a single massless particle ($\widetilde{H}_i = 0, k_C = 0$) propagating in free space, we obtain from (5) and (6) the equation

$$\widetilde{H}_4 \psi = -\sum_{\mu=1}^{4}\frac{\partial^2 \psi}{\partial x_\mu^2} - \frac{\partial^2 \psi}{\partial R^2} - \frac{3}{R}\frac{\partial \psi}{\partial R} + \frac{\overleftrightarrow{M^2}}{\hbar^2 R^2}\psi = 0. \qquad (9)$$

This equation applies to massless spinning elementary particles such as massless quarks and photons. However, equation (9) does not apply to neutrinos if they have a mass. We cannot exclude this possibility if neutrino oscillations really exist. The total wave function $\psi$ need not to be a scalar function. It can also be a vector, as it is the case for electromagnetic waves. However, the wave function $\psi$ is not a four-component spinor. Equation (9) has relevant solutions, which factorize in the form

$$\psi(x_{t\mu}, x_{r\mu}) = \psi_t(x_\mu)\, \psi_r(R, \alpha, \theta, \phi). \qquad (10)$$

Owing to the separation of the translational variables from the rotational variables, the motion of the particle in Minkowski space does not affect its 4D rotation and vice versa.

In order to obtain massive particles, we assume that they are composed of massless sub-particles. Free massless particles propagate with the velocity of light and the coupling between their transversal and intrinsic rotational motion is zero. As a result, $\psi_t$ and $\psi_r$ are solutions of the independent equations

$$\widetilde{H}_t \psi_t = -\Delta_t^{(4)} \psi_t = -\sum_{\mu=1}^{3} \frac{\partial^2 \psi_t}{\partial x_\mu^2} + \frac{\partial^2 \psi_t}{c^2 \partial t^2} = 0, \tag{11}$$

$$\widetilde{H}_r \psi_r = -\Delta_t^{(4)} \psi_r = -\frac{\partial^2 \psi_r}{\partial R^2} - \frac{3}{R}\frac{\partial \psi_r}{\partial R} + \frac{\overline{M}^2}{\hbar^2 R^2}\psi_r = 0. \tag{12}$$

Equation (11) is the standard equation for the propagation of a wave in free space, whereas equation (12) accounts for 4D-rotation of the superposed eddy wave. Contrary to the translational solutions, the rotational solutions $\psi_r$ are quantized representing four-dimensional angular-momentum eigenstates.

### 3. Pseudo hyper-spherical coordinates

In order to find the 4D rotational states in the space-like region of the Minkowski space, we introduce the pseudo hyper-spherical coordinates $R, i\alpha, \theta, \phi$, where $\alpha$ is a pseudo angle defining space-time rotations, which represent Lorentz transformations. By choosing the time-like coordinate ($x_{r4} = ict_r = ic(t - t_0)$) as the reference axis, we define unambiguously the orientation of the pseudo hyper-spherical coordinate system in the 4D space. Such a peculiar reference axis does not exist in the three-dimensional free space. The hyper-spherical coordinates are relative coordinates with respect to the origin $\vec{R}_0$ with coordinates $x_{\nu 0}, \nu = 1, 2, 3, 4$. Without loss of generality, we can put $\vec{R}_0 = 0$. Within the frame of the Minkowski metric, the relations between the Cartesian coordinates $x_\nu$ and the hyper-spherical coordinates $R, i\alpha, \theta, \phi$ in the case $R \geq 0$ are

$$x_{r4} = x_4 - x_{40} = iR \sinh \alpha,$$

$$x_{r3} = x_3 - x_{30} = R \cosh \alpha \cos \theta,$$

$$x_{r2} = x_2 - x_{20} = R \cosh \alpha \sin \theta \sin \phi,$$

$$x_{r1} = x_1 - x_{10} = R \cosh \alpha \sin \theta \cos \phi. \tag{13}$$

Here $\theta$ and $\phi$ are the polar angle and the azimuth angle, respectively. The pseudo angle $\alpha$ describes a time-like rotation of the 4D-spin within the hyper-plane embedding the three-dimensional radius vector $\vec{r}$ and the time-like $x_{r4}$-axis. The range of the space-like hyper-spherical coordinates (13) is

$$0 \leq R \leq \infty, \quad -\infty \leq \alpha \leq \infty, \quad 0 \leq \theta \leq \pi, \quad 0 \leq \phi \leq 2\pi. \tag{14}$$

The pseudo angle $\alpha$ is positive for $t_r = t_1 - t_2 > 0$ and negative in the case $t_1 - t_2 < 0$. In the first case the point 1 is going forward in time with respect to the point 2, whereas in the second case the point 1 is going "backward in time" with respect to the point 2. We separate the two cases by defining the space-time rotation for positive pseudo angles $\alpha$ as right handed and as left handed for negative values of $\alpha$.

The three-dimensional radius $r = R \cosh \alpha$ is the 3D component of the 4D radius vector $\vec{R}$. We show the physical relevance of the imaginary angle $i\alpha$ by considering an infinitesimal space-time rotation in Minkowski space. Assuming a differential angular change $d\alpha$ by keeping the radius $R$ constant, we obtain

$$-idx_{r4} = cdt_r = R \cosh \alpha \, d\alpha, \quad dr_r = R \sinh \alpha \, d\alpha, \quad \frac{idr_r}{dx_{r4}} = \frac{v_r}{c} = \beta = \tanh \alpha. \tag{15}$$

Here $v_r$ is the velocity of the particle in the three-dimensional space. By keeping $R$ constant, the space-time rotation obtained by changing the pseudo-angle $\alpha$ from $\alpha$ to $\alpha + \alpha_0$ represents a Lorentz transformation, which describes the transition from one inertial system to another inertial system. The angle $\phi$ accounts for a rotation within the three-dimensional space-like hyper-plane of the 4D-space perpendicular to the time-like $x_{r4}$-axis. Therefore, we assign this rotation to the spin of the particle. However, a change of $\phi$ by $2\pi$ does not lead to the initial state in the case of spin-1/2 particles.

By going from three to four dimensions, we cannot describe rotations in 4D-space by axial vectors but must employ an anti-symmetric second-rank tensor consisting of 12 components. This behavior results from the property of the four-dimensional cubic, which has 12 surfaces, twice as many as the three-dimensional cube. Each component of the 4D angular momentum is its projection onto the normal of the surface element $d\sigma_{\mu\nu}$. Because the normal direction of conjugate top and bottom surfaces are opposite to each other, only six independent components exist for the four-dimensional angular momentum tensor. We define the intrinsic 4D angular momentum of the elementary particles by three quantum numbers, each of which relates to one of the three angles $\alpha, \theta, \phi$. In order to obtain bound rotational states the rotational 4D system must be closed, which is the case if the four-dimensional radial vector $\vec{R}$ is *space-like*:

$$\sum_{\nu=1}^{4} x_{r\nu}^2 = R^2(\cosh^2 \alpha - \sinh^2 \alpha) = R^2 \geq 0 \ . \tag{16}$$

By employing the relations (13) relating the Cartesian coordinates with the hyper-spherical coordinates, we obtain the four-dimensional differential volume element $dV^{(4)} = dV_r^{(4)}$ in hyper-spherical coordinates by evaluating the four-dimensional functional determinant $D_f$, giving

$$dV_r^{(4)} = -if_c \prod_{\mu=1}^{4} dx_{r\mu} = -if_c D_f dR d\alpha d\theta d\phi = R^3 \cosh^2 \alpha \, dRd\Omega^{(4)}, \tag{17}$$

$$f_c = \sqrt{1 - \frac{v^2}{c^2}} = \frac{1}{\gamma} = \frac{1}{\cosh \alpha}, \quad D_f = iR^3 \cosh^2 \alpha \sin \theta, \tag{18}$$

$$d\Omega^{(4)} = f_c \sin \theta \, d\alpha d\theta d\phi = \frac{d\alpha}{\cosh \alpha} \sin \theta \, d\theta d\phi \tag{19}$$

We shall demonstrate the need for incorporating the FitzGerald-Lorentz contraction factor $f_c$ in the context of the orthogonality relation for the eigenfunctions $\psi_\alpha(\alpha)$. The contraction factor (18) guarantees that total 4D solid angle $\Omega_t^{(4)} = 4\pi^2$ stays finite by transforming the time-like polar angle $\alpha$ into the angle

$$\vartheta = \int_{-\infty}^{\alpha} \frac{d\alpha}{\cosh \alpha} = 2\tan^{-1} e^\alpha, \quad \vartheta = \begin{cases} \pi & \alpha = \infty \\ 0 & \alpha = -\infty \end{cases}. \tag{20}$$

The transformation (20) shows that the space-time angle $\alpha$ has the property of a pseudo-polar angle because $\vartheta$ extends over the same range as the spatial polar angle $\theta$.

Physically relevant solutions $\psi_{r\Omega}$ of the rotational wave equation (12) are those, which one can normalize. These functions form a set of orthonormal eigenfunctions $\psi_{\Omega n} = \psi_{\Omega n}(\alpha, \theta, \phi)$, which satisfy the orthogonal relation

$$\langle \psi_{\Omega n} \psi_{\Omega m}^\dagger \rangle = \iint \psi_{\Omega n} \psi_{\Omega m}^\dagger \cosh^2 \alpha \, d\Omega^{(4)} =$$

$$\int_0^\infty \int_{-\infty}^\infty \int_0^\pi \int_0^{2\pi} \psi_{\Omega n} \psi_{\Omega m}^\dagger \cosh \alpha \sin \theta \, d\alpha d\theta d\phi = \delta_{nm}, \quad n, m = 1, 2, \ldots, \tag{21}$$

$$\psi_{\Omega n} = \psi_{\Omega n}(\alpha, \theta, \phi) = \psi_\alpha(\alpha) \psi_\theta(\theta) \psi_\phi(\phi). \tag{22}$$

The Kronecker symbol $\delta_{mn}$ is 1 for $m = n$ and it is 0 in the case $m \neq n$. Within the frame of our 4D-treatment the adjoint wave function $\psi_{\Omega m}^{\dagger}$ is rotational wave function of the antiparticle whose 4D helicity is opposite to that of the particle. Accordingly, the normalization condition $\delta_{nn} = 1$ implies that an antiparticle with opposite 4D angular momentum must exist for each particle.

## 4. Four-dimensional angular momentum

The three-dimensional angular momentum is an axial vector consisting of three components whereas the 4D angular momentum is an anti-symmetric second-rank tensor consisting of six components. By introducing Cartesian coordinates, the components $M_{\mu\nu} = -M_{\nu\mu}$ of the 4D angular momentum operator $\overleftrightarrow{M} = i\hbar\overleftrightarrow{\Omega}$ are

$$M_{\mu\nu} = p_\mu x_\nu - p_\nu x_\mu = \frac{\hbar}{i}\left(x_\nu \frac{\partial}{\partial x_\mu} - x_\mu \frac{\partial}{\partial x_\nu}\right) = i\hbar\Omega_{\mu\nu}, \quad \mu,\nu = 1,2,3,4. \quad (23)$$

The normals of each of the six conjugate top and bottom surfaces of the 4D cube have opposite directions. Therefore, only six independent components $M_{\mu\nu} = -M_{\nu\mu}$ exist, three of which $(M_{12}, M_{23}, M_{31})$ are imaginary and three are real $(M_{14}, M_{24}, M_{34})$. Instead of the angular momentum tensor $\overleftrightarrow{M}$, it is advantageous to employ the dimension-less anti-symmetric angular tensor operator $\overleftrightarrow{\Omega} = -i\overleftrightarrow{M}/\hbar$ with components $\Omega_{\mu\nu} = -\Omega_{\nu\mu}$. The space-like components $\Omega_{12}, \Omega_{23}, \Omega_{31}$ of the angular momentum operator $\overleftrightarrow{\Omega}$ are real whereas the time-like components $\Omega_{14}, \Omega_{24}, \Omega_{34}$ are imaginary, as it is the case for the components $F_{\mu\nu}$ of the electromagnetic field tensor. The operators $\Omega_{\mu\nu}$ satisfy the commutation relation

$$\Omega_{\mu\lambda}\Omega_{\nu\lambda} - \Omega_{\nu\lambda}\Omega_{\mu\lambda} = \Omega_{\mu\nu}. \quad (24)$$

The operators $2i\Omega_{\mu\nu}$ are the equivalents of the Dirac spinor operators $\sigma_{\mu\nu} = -i\gamma_\mu\gamma_\nu$ because both describe intrinsic rotations (spin) and satisfy the same commutation relation (24). However, within the frame of the Dirac algebra the spinor operators are 4by4 matrices acting on the four-component spinor whereas the operators $\Omega_{\mu\nu}$ are differential operators acting on the angular wave function $\psi_\Omega = \psi_\Omega(\alpha, \theta, \phi)$. Replacing the Dirac spin operators by the differential operators $\Omega_{\mu\nu}$ corresponds to the transition from the Heisenberg matrix formalism to the Schrödinger representation of classical quantum mechanics.

In order to find the angular eigenfunctions $\psi_\Omega = \psi_\Omega(\alpha, \theta, \phi)$ and the eigenvalues of the square $\overleftrightarrow{\Omega}^2$ of the angular tensor operator, we must express the operator $\overleftrightarrow{\Omega}^2$ in hyper-spherical angular coordinates $\alpha, \theta, \phi$. We derive this representation by rewriting the components (23) of the angular tensor operators $\overleftrightarrow{\Omega}$ in hyper-spherical coordinates. As a result, we find that the space-like components $\vec{\Omega}_m = \vec{e}_x \Omega_{mx} + \vec{e}_y \Omega_{my} + \vec{e}_z \Omega_{mz}$ of the angular tensor operator coincide with the angular momentum operator of standard quantum mechanics [8], given by

$$\Omega_{23} = \Omega_{mx} = \sin\phi \frac{\partial}{\partial \theta} + \cot\theta \cos\phi \frac{\partial}{\partial \phi}, \tag{25}$$

$$\Omega_{31} = \Omega_{my} = -\cos\phi \frac{\partial}{\partial \theta} + \cot\theta \sin\phi \frac{\partial}{\partial \phi}, \tag{26}$$

$$\Omega_{12} = \Omega_{mz} = -\frac{\partial}{\partial \phi}. \tag{27}$$

The three components of the time-like component $\vec{\Omega}_e = \vec{e}_x \Omega_{ex} + \vec{e}_y \Omega_{ey} + \vec{e}_z \Omega_{ez}$ of the angular tensor operator are:

$$i\Omega_{14} = \Omega_{ex} = \cos\phi \left(\sin\theta \frac{\partial}{\partial \alpha} + \tanh\alpha \cos\theta \frac{\partial}{\partial \theta}\right) - \frac{\sin\phi}{\sin\theta} \tanh\alpha \frac{\partial}{\partial \phi}, \tag{28}$$

$$i\Omega_{24} = \Omega_{ey} = \sin\phi \left(\sin\theta \frac{\partial}{\partial \alpha} + \tanh\alpha \cos\theta \frac{\partial}{\partial \theta}\right) + \frac{\cos\phi}{\sin\theta} \tanh\alpha \frac{\partial}{\partial \phi}, \tag{29}$$

$$i\Omega_{34} = \Omega_{ez} = \cos\theta \frac{\partial}{\partial \alpha} - \tanh\alpha \sin\theta \frac{\partial}{\partial \theta}. \tag{30}$$

The space-like component $\vec{\Omega}_s$ of the angular tensor operator accounts for an intrinsic rotation (spin) about an axis in the three-dimensional subspace whereas the time-like component $\vec{\Omega}_t$ accounts for an imaginary rotation within a hyperplane, which embeds the time-like axis.

The physical properties of the angular vector operators $\vec{\Omega}_s$ and $\vec{\Omega}_t$ are best illustrated by introducing the spherical unit vectors $\vec{e}_r, \vec{e}_\theta, \vec{e}_\phi$, giving

$$\vec{\Omega}_m = -\vec{e}_\phi \frac{\partial}{\partial \theta} + \vec{e}_\theta \frac{\partial}{\sin\theta \partial \phi}, \tag{31}$$

$$\vec{\Omega}_e = \vec{e}_r \frac{\partial}{\partial \alpha} + \tanh\alpha \left(\vec{e}_r \times \vec{\Omega}_s\right) = \vec{e}_r \frac{\partial}{\partial \alpha} + \tanh\alpha \left(\vec{e}_\theta \frac{\partial}{\partial \theta} + \frac{\vec{e}_\phi}{\sin\theta} \frac{\partial}{\partial \phi}\right). \tag{32}$$

The radial component $\vec{e}_r \partial/\partial\alpha$ of the time-like operator $\vec{\Omega}_e$ accounts for an acceleration in radial direction. This operator has the property of an electric monopole and couples with an equivalent time-like quantity such as the electric field of an atomic nucleus. However, the direct coupling of the imaginary time-like component of the electromagnetic field tensor with the real time-like component of the angular momentum tensor does not yield a real quantity, as it must be for the energy. Moreover, $d\alpha$ represents an infinitesimal change of the imaginary rotation, which depends on the velocity $v = \beta c = \tanh\alpha$. For example, the orbital motion and transitions of the electron from its ground state to an excited state and vice versa require an acceleration of the electron in radial direction. The operators $\partial/\partial\theta$ and $\partial/\partial\phi$ account for real rotations in the three-dimensional laboratory space. We can interpret the term $\tanh\alpha\,(\vec{e}_r \times \vec{\Omega}_m)$ of the operator (32) by means of a standard Lorentz transformation for the field of a magnetic dipole from one inertial system to another. By considering the relation $\tanh\alpha = \beta = v/c$, we realize that the time-like operator $(\vec{e}_r \times \vec{\Omega}_m)\tanh\alpha$ is the equivalent of the electric field induced by a moving magnetic dipole with relative velocity $v/c$. This equivalence shows that we can conceive $\vec{\Omega}_m$ as the magnetic-field operator and $\vec{\Omega}_e$ as the electric-field operator. The vector operators $\vec{\Omega}_m$ and $\vec{\Omega}_e$ are orthogonal to each other and satisfy the relation

$$\vec{\Omega}_m \cdot \vec{\Omega}_e = -\vec{\Omega}_e \cdot \vec{\Omega}_m = i(\Omega_{12}\Omega_{34} - \Omega_{13}\Omega_{24} + \Omega_{14}\Omega_{23}) = 0. \tag{33}$$

This relation shows that of the vector operators $\vec{\Omega}_m$ and $\vec{\Omega}_e$ commute for their scalar product whereas their components do not. We obtain the orthogonality relation for the electric field strength $\vec{E}$ and the magnetic field strength $\vec{B}$ of electromagnetic waves by substituting the components $F_{\mu\nu} = -F_{\nu\mu}$ of the electromagnetic field tensor for $\Omega_{\mu\nu}$ in (33), giving

$$F_{12}F_{34} - F_{13}F_{24} + F_{14}F_{23} = F_{24}F_{13} - F_{23}F_{14} + F_{34}F_{21} = \frac{i}{c}\vec{E}\vec{B} = 0. \tag{34}$$

Owing to this equivalence, we propose that the imaginary time-like operator $\vec{\Omega}_e$ represents the electrical component of the angular tensor operator $\overleftrightarrow{\Omega}$ and the operator $\vec{\Omega}_m$ the magnetic component. The similarity between the properties of the components of the electromagnetic field tensor and those of the four-dimensional angular tensor operator. Therefore, we assume that they represent electromagnetic operators producing the electromagnetic field of the particle. Owing to this behavior, our approach is equivalent to quantum electrodynamics.

We obtain the representations of the operators $\vec{\Omega}_m^2$, $\vec{\Omega}_e^2$, and $i\vec{\Omega}_m \times \vec{\Omega}_e$ in hyper-spherical coordinates by employing the relations (25) - (30), giving

$$\vec{\Omega}_m^2 = \frac{1}{\sin\theta}\frac{\partial}{\partial\theta}\left(\sin\theta\frac{\partial}{\partial\theta}\right) + \frac{1}{\sin^2\theta}\frac{\partial^2}{\partial\phi^2}, \tag{35}$$

$$\sum_{\mu=1}^{3}\Omega_{\mu 4}^2 = -\vec{\Omega}_e^2 = -\frac{\partial^2}{\partial\alpha^2} - 2\tanh\alpha\frac{\partial}{\partial\alpha} - \tanh^2\alpha \cdot \vec{\Omega}_s^2, \tag{36}$$

$$\vec{\Omega}_f = i\vec{\Omega}_m \times \vec{\Omega}_e = -i\vec{\Omega}_e \times \vec{\Omega}_m = i\vec{e}_r\tanh\alpha\,\vec{\Omega}_m^2 + i\left(\vec{e}_\theta\frac{\partial}{\partial\theta} + \frac{\vec{e}_\phi}{\sin\theta}\frac{\partial}{\partial\phi}\right)\left(\tanh\alpha - \frac{\partial}{\partial\alpha}\right). \tag{37}$$

The structure of the space-like operator (35) differs from that of the time-like operator (36) whereas the equivalent operators for the magnetic field strength $\vec{B}$ and the electric field strength $\vec{E}$ of free electromagnetic waves are identical. The time-like operator $\vec{\Omega}_f$ represents the angular-momentum-flux operator or energy-flux operator, respectively, which represents the quantum-mechanical equivalent to the Poynting vector $\vec{P} = \vec{E} \times \vec{B}/c$ of classical electrodynamics with the exception that the electric and magnetic field strengths are physical quantities and not operators. In order that our interpretation of the operator (37) is realistic, its expectation value must be non-zero for transitions from one quantum state to another and zero else.

By considering the orthogonality relation (33), we eventually obtain the eigenvalues $J$ of the square $\overleftrightarrow{M}^2$ of the angular tensor operator $\overleftrightarrow{M}$ the equation

$$\overleftrightarrow{M}^2\psi_\Omega = -\hbar^2\overleftrightarrow{\Omega}^2\psi_\Omega = \hbar^2 J(J+1)\psi_\Omega, \quad \psi_\Omega = \psi_\Omega(\alpha,\theta,\phi),$$

$$\overleftrightarrow{\Omega}^2 = \left(\vec{\Omega}_m - i\vec{\Omega}_e\right)^2 = \vec{\Omega}_m^2 - \vec{\Omega}_e^2 = \sum_{\mu<\nu}\Omega_{\mu\nu}^2 =$$

$$-\frac{\partial^2}{\partial\alpha^2} - 2\tanh\alpha\frac{\partial}{\partial\alpha} + \frac{1}{\cosh^2\alpha}\left\{\frac{1}{\sin\theta}\frac{\partial}{\partial\theta}\left(\sin\theta\frac{\partial}{\partial\theta}\right) + \frac{1}{\sin^2\theta}\frac{\partial^2}{\partial\phi^2}\right\}. \tag{38}$$

The Lorentz-invariant operator $\overleftrightarrow{\Omega}^2$ accounts for the intrinsic rotations of the particle, in particular the term $\partial^2/\partial\alpha^2$ for an acceleration. Owing to the representation (38), we conceive the particle as a four-dimensional spinning top propagating in the 4D-space with angular-momentum eigenvalue $J$. In order that the 4D rotation is close, the eigenvalue J must be real. The Lorentz-invariant equivalent of the operator $\overleftrightarrow{\Omega}^2 = \sum_{\mu<\nu}\Omega_{\mu\nu}^2$ for electromagnetic waves is $\overleftrightarrow{F}^2 = \sum_{\mu<\nu}F_{\mu\nu}^2 = \vec{B}^2 - \vec{E}^2/c^2 = 0$. However, a distinct difference exists regarding the properties of the components $F_{\mu\nu}$ and $\Omega_{\mu\nu}$. The components $F_{\mu\nu}$ of the electromagnetic field tensor $\overleftrightarrow{F}$ are measurable physical quantities, whereas the components $\Omega_{\mu\nu}$ of the tensor $\overleftrightarrow{\Omega}$ are angular operators.

The rotational eigenfunctions of the elementary particles are solutions of the rotational Hamiltonian (13). Because this Hamiltonian is real, it describes a static state ($\partial \psi / \partial \tau = 0$), whose rotational wave function $\psi_r$ satisfies the equation

$$-\widetilde{H}_{4r}\psi_r = \Delta_r^{(4)}\psi_r = \frac{\partial^2 \psi_r}{\partial R^2} + \frac{3}{R}\frac{\partial \psi_r}{\partial R} - \frac{1}{R^2}\left(\frac{\partial^2 \psi_r}{\partial \alpha^2} + 2\tanh\alpha \frac{\partial \psi_r}{\partial \alpha}\right) +$$

$$\frac{1}{R^2 \cosh^2 \alpha}\left\{\frac{1}{\sin\theta}\frac{\partial}{\partial \theta}\left(\sin\theta \frac{\partial \psi_r}{\partial \theta}\right) + \frac{1}{\sin^2\theta}\frac{\partial^2 \psi_r}{\partial \phi^2}\right\} = 0. \tag{39}$$

We solve this equation by employing Bernoulli's separation procedure yielding solutions, which factorize in the form

$$\psi_r(R, \alpha, \theta, \phi) = \psi_R(R)\psi_\Omega(\alpha, \theta, \phi) = \psi_R(R)\psi_\alpha(\alpha)\psi_{\theta\phi}(\theta, \phi). \tag{40}$$

By substituting (40) for $\psi_r$ in equation (38) and by employing the Bernoulli separation procedure, we obtain the decoupled equations

$$\frac{d^2 \psi_R}{dR^2} + \frac{3}{R}\frac{d\psi_R}{dR} + \frac{1-\mu^2}{R^2}\psi_R = 0, \quad 1 - \mu^2 = J(J+1) \geq 0, \tag{41}$$

$$\frac{d^2 \psi_\alpha}{d\alpha^2} + 2\tanh\alpha + (1 - \mu^2)\psi_\alpha + \frac{j(j+1)}{\cosh^2 \alpha}\psi_\alpha = 0, \tag{42}$$

$$\frac{1}{\sin\theta}\frac{\partial}{\partial \theta}\left(\sin\theta \frac{\partial \psi_{\theta\phi}}{\partial \theta}\right) + \frac{1}{\sin^2\theta}\frac{\partial^2 \psi_{\theta\phi}}{\partial \phi^2} + j(j+1)\psi_{\theta\phi} = 0. \tag{43}$$

The separation constants $j$ and $\mu$ are eigenvalues. It follows from the second relation of (41) that the angular momentum eigenvalue $J$ is only real if $\mu$ is in the range $0 \leq |\mu| \leq 1$. As a result, we only obtain one half-integer eigenvalue $J = 1/2$ for $\mu = \pm 1/2$ and one integer eigenvalue $J = 0$ in the case $\mu = \pm 1$. In the first case, the 4D angular momentum is space-like defining a closed system, whereas it is light-like in the latter case representing the 4D angular momentum of radiation quanta such as photons. The eigenvalue $\mu$ of the equation (42) represents the "electric-spin" quantum number of the 4D angular momentum. We assign the minus sign to the negative charge and the plus sign to the positive charge.

It readily follows from (35), (38), (42), and (43) that the 4D angular eigenfunctions $\psi_\Omega = \psi_\alpha(\alpha)\psi_{\theta\phi}(\theta, \phi)$ satisfy the equations

$$\vec{\Omega}^2\psi_\Omega = (\vec{\Omega}_m^2 - \vec{\Omega}_e^2)\psi_\Omega = (1-\mu^2)\psi_\Omega, \qquad \vec{\Omega}_m^2\psi_\Omega = -j(j+1)\psi_\Omega. \tag{44}$$

The quantum number *j* of the familiar three-dimensional angular equation (43) has integer and half-integer values defining the space-like component of the 4D angular momentum. In the case of an integer quantum number $j = l$, the solutions of (43) are the spherical harmonics

$$\psi_{\theta\phi} = Y_l^m(\theta,\phi), \quad l = 1, 2, \ldots, m = -l, -l+1, \ldots, l-1, l. \tag{45}$$

The azimuth quantum number *m* has positive and negative values depending on the helicity of the quantum state. We obtain the time-like wave function $\psi_\alpha(\alpha)$ by substituting the modified wave function $\Psi_\alpha(\alpha) = \psi_\alpha(\alpha)\cosh\alpha$ for $\psi_\alpha$ in (42), giving

$$\frac{d^2\Psi_\alpha}{d^2\alpha} - \mu^2\Psi_\alpha + \frac{j(j+1)}{\cosh^2\alpha}\Psi_\alpha = 0. \tag{46}$$

Normalizable solutions of this equation exist for $0 \le \mu^2 \le 1$ in the case that *j* or $j - \mu = n$ are integers or that they are zero. We attribute the light-like eigenvalue $\mu = \pm 1$ to mass-less particles, e.g. photons. The time-like eigenvalues $\mu = \pm 1/2$ characterize charged spin-1/2 particles for integer values $j = l_s = 0, 1, \ldots$ and neutral spin-1/2 particles for half-integer values $j = j_s = \frac{1}{2}, \frac{3}{2}, \ldots$ .

Particles whose eigenvalues satisfy the relation $j - |\mu| = n = 0, 1, \ldots$ are neutral. Their time-like angular wave function does not depend on the sign of the eigenvalue $\mu$:

$$\psi_{\alpha n}^{(\mu)}(\alpha) = \frac{\Psi_{\alpha n}^{(\mu)}(\alpha)}{\cosh\alpha} = c_{n\mu}\frac{C_n^{(\mu)}(\tanh\alpha)}{\cosh^{1+\mu}\alpha}, \quad \mu = |\mu| \ge 1/2, \; n = 0, 1, 2, \ldots, \tag{47}$$

$$C_n^{(\mu)}(\tanh\alpha) = \frac{1}{\Gamma(\mu)}\sum_{m=0}^{[\frac{n}{2}]}(-1)^m\frac{\Gamma(\mu+n-m)}{m!(n-2m)!}(2\tanh\alpha)^{n-2m}. \tag{48}$$

The functions $C_n^{(\mu)}(\tanh\alpha)$ are the ultra-spherical (Gegenbauer) polynomials [9]. The upper limit [*n*/2] of the summation is the integer value of the quotient $n/2$. Because the solutions (47) do not depend on the sign of the eigenvalue $\mu$, we must assign them to neutral particles.

The solutions $\psi_{\alpha n}^{(\mu)}(\alpha)$ form a complete set of orthogonal eigenfunctions, which satisfy the orthogonality relation

$$\int_{-\infty}^{\infty} \psi_{\alpha n}^{(\mu)}(\alpha)\psi_{\alpha m}^{(\mu)}(\alpha) \cosh \alpha \, d\alpha = \int_{-\infty}^{\infty} \Psi_{\alpha n}^{(\mu)}(\alpha)\Psi_{\alpha m}^{(\mu)}(\alpha) \frac{d\alpha}{\cosh \alpha} =$$

$$\frac{4^\mu}{2\pi} \frac{n!(n+\mu)[\Gamma(\mu)]^2}{\Gamma(2\mu+n)} \int_{-1}^{1} C_n^{(\mu)}(u) C_m^{(\mu)}(u)(1-u^2)^{J-1/2} du = \delta_{mn}, \quad u = \tanh \alpha. \tag{49}$$

By employing the normalization relation for the ultra-spherical polynomials $C_n^{(\mu)}(u)$ [9] and considering (49), we eventually obtain the normalized eigenfunctions

$$\psi_{\alpha n}^{(\mu)}(\alpha) = \frac{\Psi_{\alpha n}^{(\mu)}(\alpha)}{\cosh \alpha} = 2^J \Gamma(\mu) \sqrt{\frac{n!(n+\mu)}{2\pi \Gamma(2\mu+n)}} \frac{C_n^{(\mu)}(\tanh \alpha)}{\cosh^{1+\mu} \alpha} . \tag{50}$$

Solutions having negative quantum number *n* do not exist. By taking into account the relation (45), we obtain the light-like 4D angular wave function $\psi_\Omega(\alpha, \theta, \phi) = \psi_\alpha(\alpha)\psi_{\theta\phi}(\theta, \phi)$ for integer quantum number $j = l > 0$ and $\mu = \pm 1$ in the form

$$\psi_\Omega(\alpha, \theta, \phi) = 2^l \frac{C_{l-1}^{(1)}(\tanh \alpha)}{\sqrt{2\pi} \cosh^2 \alpha} Y_l^m(\theta, \phi), \quad l = 1, 2, \dots . \tag{51}$$

By putting $\mu = \pm 1$ in the equation (41), we obtain the solution $\psi_R = \psi_{R0} = C_{R0}/R_0^2$, which represents the radial term of the total rotational wave function $\psi_r = \psi_R \psi_\alpha(\alpha)\psi_{\theta\phi}(\theta, \phi)$. The elementary radius $R_0$ is a characteristic elementary length of the light-like 4D angular momentum ($J = 0$).

The eigenvalues $\mu = |\mu|$ and *j* of neutral particles are both either integers or half-integers. Moreover, the associated eigenfunctions do not depend on the sign of $\mu$. However, the 4D angular wave function of charged spin-1/2 particles must depend on the sign of the time-like eigenvalue $\mu$, which defines the sign of the charge. We obtain the eigenfunctions of neutral and charged spin-1/2 particles by introducing the modified wave functions

$$\Psi_\Omega = \psi_\Omega \cosh \alpha \sqrt{\sin \theta} = \Psi_\alpha(\alpha)\Psi_{\theta\phi}(\theta, \phi), \tag{52}$$

$$\Psi_\alpha(\alpha) = \psi_\alpha(\alpha) \cosh \alpha, \quad \Psi_{\theta\phi}(\theta, \phi) = \psi_{\theta\phi}\sqrt{\sin \theta} . \tag{53}$$

By employing the relations (53), we rewrite equations (42) and (43) in the form

$$\frac{d^2\Psi_\alpha}{d^2\alpha} - \mu^2 \Psi_\alpha + \frac{j(j+1)}{\cosh^2 \alpha} \Psi_\alpha = 0, \quad \mu = \pm\tfrac{1}{2},\ j = 0, \tfrac{1}{2}, 1, \ldots, \tag{54}$$

$$\frac{\partial^2 \Psi_{\theta\phi}}{\partial^2 \theta} + \frac{1}{\sin^2 \theta}\left(\tfrac{1}{4}\Psi_{\theta\phi} + \frac{\partial^2 \Psi_{\theta\phi}}{\partial\phi^2}\right) + \left(j + \tfrac{1}{2}.\right)^2 \Psi_{\theta\phi} = 0. \tag{55}$$

Equation (55) has for each quantum number $j = 0, \tfrac{1}{2}, 1, \tfrac{3}{2}, \ldots$, the four linearly independent solutions

$$\Psi_{\theta\phi} = \tfrac{1}{\pi} e^{is\phi} e^{ij_s\theta}, \quad s = \pm\tfrac{1}{2}, \quad j_s = \pm\left(j + \tfrac{1}{2}\right). \tag{56}$$

Because the eigenvalues $j_s$ are associated with the polar angle $\theta$, they define the precession state of the spin. We separate the solutions (56) in two categories $\Psi^{(n)}_{\theta\phi}$ and $\Psi^{(c)}_{\theta\phi}$ depending on the eigenvalues $j_s$:

$$\Psi^{(n)}_{\theta\phi} = \tfrac{1}{\pi} e^{is\phi} e^{ij_s\theta},\ j_s = \pm 1, \pm 2, \ldots, \quad \Psi^{(c)}_{\theta\phi} = \tfrac{1}{\pi} e^{is\phi} e^{ij_s\theta},\ j_s = \pm\tfrac{1}{2}, \pm\tfrac{3}{2}, \ldots. \tag{57}$$

We assign the solutions with half-integer eigenvalues $j_s$ to charged particles and the solutions with integer $j_s$ to neutral spin-1/2 particles, such as neutrinos. The eigenfunctions $\Psi^{(c)}_{\theta\phi}$ with half-integer $j_s$ do not form a set of orthogonal functions whereas the eigenfunctions $\Psi^{(n)}_{\theta\phi}$ with different integer quantum numbers $j_s$ are orthogonal to each other. We face the same situation for the time-like rotations.

The appearance of the half-angle $s\phi = \pm\phi/2$ has the peculiar consequence that for a $\phi = 360°$ rotation the wave function $\psi_\phi = e^{is\phi}$ differs from its initial value by a minus sign. Accordingly, we need a $\phi = 720°$ rotation to get back to the original state. Two groups have demonstrated experimentally the physical reality of the predicted minus sign for the $360°$ rotation [10, 11]. Therefore, we cannot exclude the physically relevant solutions (56, 57). We can conceive the puzzling experimental result by considering that a complete rotation within a hyper-plane of the 4D space corresponds in the there-dimensional space to the total solid angle, which is $4\pi$. Although $\psi_{\theta\phi} = \Psi_{\theta\phi}/\sqrt{\sin\theta}$ is weakly singular at $\theta = 0$ and $\theta = \pi$, it can be normalized because the differential probability density $|\psi_{\theta\phi}|^2 \sin\theta d\theta = |\Psi_{\theta\phi}|^2 d\theta$ stays finite in the entire region $0 \leq \theta \leq \pi$. By considering that $\sin\theta$ is a metric coefficient, we propose that $\Psi_\Omega = \Psi_\alpha(\alpha)\Psi_{\theta\phi}(\theta,\phi)$ is the relevant angular wave function, which is free of singularities.

The eigenfunctions $\psi_\Omega = \Psi_\Omega/\cosh\alpha\sqrt{\sin\theta}$ satisfy the normalization condition (19)

$$\int \psi_\Omega \psi_\Omega^\dagger f_c d\Omega^{(4)} = \int_{-\infty}^\infty \int_0^\pi \int_0^{2\pi} \Psi_\Omega \Psi_\Omega^\dagger \frac{d\alpha}{\cosh\alpha} d\theta d\phi = 1, \tag{58}$$

By substituting $d\vartheta$ for $d\alpha/\cosh\alpha$ in the last integral of (58), we find that the modified wave function $\Psi_\Omega$ is physically more relevant than $\psi_\Omega$ because the resulting integrand is the four-dimensional angular-momentum density $\Psi_\Omega \Psi_\Omega^\dagger$, which does not depend explicitly on the metric coefficients $\sin\theta$ and $\cosh\alpha$; $\psi_\Omega^\dagger$ and $\Psi_\Omega^\dagger$ are the adjoint wave functions.

For integer values $j = l_s$ equation (46) has the linearly independent solutions

$$\Psi_\alpha(\alpha) = \Psi_{\alpha l_s}^{(\mu)}(\alpha) = \psi_{\alpha l_s}^{(\mu)} \cosh\alpha = C_{l_s}^{(\mu)} \cosh^{l_s+1}\alpha \left(\frac{1}{\cosh\alpha}\frac{d}{d\alpha}\right)^{l_s} \frac{e^{\mu\alpha}}{\cosh\alpha},$$

$$-\infty \leq \alpha \leq \infty, \qquad l_s = |j_s| - \frac{1}{2} = 0, 1, 2, \ldots. \tag{59}$$

The parameter $\mu$ can adopt arbitrary values. The wave function $\Psi_{\alpha l_s}^{(\mu)}$ is concentrated in the region $\alpha > 0$ for $\mu > 0$ and confined to the opposite region $\alpha < 0$ in the case $\mu < 0$. Physically relevant values for $\mu$ are $\mu = \pm 1$ and $\mu = \pm 1/2$. For $\mu = \pm 1/2$ the wave functions are

$$\psi_{\alpha l_s}^{(\pm 1/2)}(\alpha) = C_{\alpha l_s}^{(\pm 1/2)} \cosh^{l_s}\alpha \left(\frac{1}{\cosh\alpha}\frac{d}{d\alpha}\right)^{l_s} \frac{e^{\pm\frac{\alpha}{2}}}{\cosh\alpha} = e^{\pm\alpha/2} D_{l_s}^{(\pm 1/2)}(\tanh\alpha), \tag{60}$$

We assign the functions (59) to charged spin-1/2 particles and their antiparticles such that $\psi_{\alpha l_s}^{(-1/2)}(\alpha) = \psi_{\alpha l_s}^{(1/2)\dagger}(\alpha)$ is the wave function of the negative-charged particle and $\psi_{\alpha l_s}^{(1/2)}(\alpha) = \psi_{\alpha l_s}^{(-1/2)\dagger}(\alpha)$ is the wave function of its positive-charged antiparticle, regardless of their mass. The wave function of the anti-particle is the adjoint wave function of the particle and vice versa. The normalization of the wave functions shows that the normalization factors $C_{l_s}^{(+)}$ and $C_{l_s}^{(-)}$ are identical ($C_{l_s}^{(\pm)} = C_{l_s}$). The time-like angular-momentum eigenfunctions $\psi_{\alpha 0}^{(\pm 1/2)}(\alpha)$ represent the equivalents of the spin eigenfunctions $e^{\pm i\phi/2}$. The polynomials $D_{l_s}^{(\pm 1/2)}(\tanh \alpha)$ do not form a set of orthogonal eigenfunctions. Therefore, states with a fixed quantum number $l_s > 0$ are not stable apart from the ground state $l_s = 0$.

By differentiating (60) with respect to the angle $\alpha$, we obtain

$$\frac{c_{l_s}}{c_{l_s+1}} \psi_{\alpha,l_s+1}^{(\pm 1/2)}(\alpha) = \frac{\partial \psi_{\alpha,l_s}^{(\pm 1/2)}(\alpha)}{\partial \alpha} - l_s \tanh\alpha \; \psi_{\alpha,l_s}^{\left(\pm\frac{1}{2}\right)}(\alpha) = \boldsymbol{t}_{l_s} \psi_{\alpha,l_s}^{(\pm 1/2)}, \tag{61}$$

$$\boldsymbol{t}_{l_s} = \frac{\partial}{\partial \alpha} - l_s \tanh\alpha. \tag{62}$$

We readily obtain from (61) the transition operator $\boldsymbol{T}_{l_s}$ for the wave functions $\Psi_{\alpha,l_s}^{\left(\pm\frac{1}{2}\right)}(\alpha)$:

$$\boldsymbol{T}_{l_s} = \boldsymbol{t}_{l_s} - \tanh\alpha = \frac{\partial}{\partial \alpha} - (l_s + 1)\tanh\alpha. \tag{63}$$

The *transition operator* $\boldsymbol{t}_{l_s}$ rises the state $l_s$ of the wave function $\psi_{\alpha,l_s}^{(\pm 1/2)}$ to the next-higher momentum state $l_s + 1$ with wave function $\psi_{\alpha,l_s+1}^{(\pm 1/2)}(\alpha)$. This behavior proves the standard rule that transitions of the electron in an atom are not possible between states of the same angular momentum *l*. Moreover, spectroscopy shows that optical transitions occur only between *l* and $l \pm 1$ states, e.g. between *s*-states and *p*-states. Owing to the conservation of angular momentum, we suppose that the intrinsic time-like angular momentum $l_s$ of the bound electron coincides with its orbital momentum *l* ($l_s = l$). In order to define a quantum-mechanical transition we must fix the initial state and the final state. However, the particle may also remain in its initial state because the corresponding transition amplitude $w_{l_s,l_s}$ can be nonzero. The transition amplitude for the two different possibilities is

$$w_{l_s l'_s} = \langle l_s | \boldsymbol{t}_{l_s} | l'_s \rangle = \int_{-\infty}^{\infty} \psi_{\alpha,l_s}^{(\pm 1/2)\dagger}(\alpha) \boldsymbol{t}_{l_s} \psi_{\alpha,l'_s}^{(\pm 1/2)} \cosh\alpha \, d\alpha \neq 0. \tag{64}$$

The result shows that we cannot assign the eigenfunctions $\psi_{\alpha,l_s}^{(\pm 1/2)}$ to stable eigenstates, whose wave functions must be orthogonal. We demonstrate this behavior by considering the transition from the ground state $l_s = 0$ to the exited state $l_s + 1 = 1$. The normalized wave function (59) for the ground state $l_s = 0$ and the *p*-state $l_s = 1$ are

$$\psi_{\alpha 0}^{(\pm 1/2)} = \frac{1}{\sqrt{\pi}} \frac{e^{\pm\frac{\alpha}{2}}}{\cosh \alpha}, \qquad \psi_{\alpha 1}^{(\pm 1/2)} = \frac{2}{\sqrt{\pi}} \frac{e^{\pm\frac{\alpha}{2}}}{\cosh \alpha} \left(\pm \frac{1}{2} - \tanh \alpha\right). \tag{65}$$

We can interpret these wave functions as the eigenfunctions of "the time-like spin", which accounts for space-time rotations. By inserting the functions (60) into the integrand of the integral (64), we obtain for the negatively charged particle the transition amplitudes

$$w_{00} = \langle 0|t_0|0\rangle = -\frac{1}{2}, \qquad w_{10} = \langle 1|t_0|0\rangle = \frac{1}{2},$$

$$w_{01} = \langle 0|t_0|1\rangle = -\frac{1}{2}, \qquad w_{11} = \langle 1|t_0|1\rangle = -\frac{3}{2}. \tag{66}$$

The transition amplitudes (64) are characteristic quantities, which we should not mix up with transition-probability amplitudes.

We derive the eigenfunctions $\Psi_{\alpha j}^{(1/2)}(\alpha)$ of neutral spin-1/2 particles from (57) by putting $\mu = 1/2$ and by taking half-integer values for the quantum number $j$ or integer values for $l_s = j - 1/2$. Therefore, the angular wave functions of neutral spin-1/2 particles are not linear combinations of those of the positive and negative charged particles, which have integer quantum numbers $j$. Such linear combinations are physically nonrealistic because we cannot normalize them.

## 5. Photon and massless quarks (quarkinos)

Stable mass-less free particles propagate with the velocity of light and their time-like spin quantum number $J$ is either an integer (photons), a half-integer $s = \pm 1/2$ (charged quarks), or zero (neutral quarks). In the absence of external fields ($\widetilde{H}_i = 0$), the translational motion and the intrinsic rotation of these massless particles are independent from each other satisfying the equations

$$\widetilde{H}_t \psi = -\sum_{\mu=1}^{4} \frac{\partial^2 \psi}{\partial x_\mu^2} = 0, \quad \widetilde{H}_r \psi = -\sum_{\mu=1}^{4} \frac{\partial^2 \psi}{\partial x_{\mu r}^2} = 0. \tag{67}$$

All known massive particles travel with velocities smaller than the velocity of light. In this case, an inertial system exists in which the particle is at rest regardless how small its mass is. Therefore, it seems rather unlikely that free neutrinos have a mass. However, bound photons, neutrinos, and basic quarks must have a mass in order that they can be at rest in an inertial system. The time-like helicity of the photon in the round state is either $\mu = 1$ or $\mu = -1$. Its space-like helicity is zero, whereas the higher states of the photons have space-like and time-like angular-momentum properties

In quantum-electrodynamic the 4D vector potential $\vec{A}_{ph}$ represents the wave function of the photon. Within the frame of our approach, the wave function of the free photon satisfies equations (11) and (12), which apply for massless particles. The general solution of the 4D angular equation in the case $\mu = 0$ and integer quantum numbers $j$ and $m$ is

$$\psi_\Omega = C_j \frac{C_{j-1}^{(1)}(\tanh \alpha)}{R_j^2 \cosh^2 \alpha} Y_j^m(\theta, \phi), \tag{68}$$

$$j = 1, 2, \dots, \quad m = -j, -j+1, \dots, j-1, j.$$

However, because these wave functions are associated with bound states we cannot assign them to photons. We must assume that the 4D angular momentum of photon in the ground state is $\mu = 0, j = 0, s = 0$. The normalized 4D wave function $\psi_{p\Omega}^{(\pm)}$ for this state angular state is

$$\psi_{p\Omega}^{(\pm)}(\alpha) = \frac{\Psi_{p\alpha}^{(\pm)}(\alpha)}{\cosh \alpha} = \frac{1}{\sqrt{\pi}} \frac{e^{\pm \alpha}}{\cosh \alpha}. \tag{69}$$

We assign the plus sign to the photon and the minus sign to the "*antiphoton*", which has opposite time-like helicity. Since the time-like rotation accounts for the charge of the particle, the photon must have a charge in the ground state $j = 0, s = 0$. However, we know from experiments that free photons do not have a charge. In order to conceive this seemingly contradiction, we assume that free photons are in neutral angular states $j = l_s = 1$. To prove this conjecture, we consider the transition from the photon ground state $l_s = 0$ to the state $l_s = 1$ by means of the transition operator (63) acting on the ground state $\psi_{\alpha 0}^{(\pm)}$, giving

$$\psi_{\alpha 1}^{(\pm)}(\alpha) = \frac{d\psi_{\alpha 0}^{(\pm)}}{d\alpha} = \frac{d}{d\alpha}\left(\frac{e^{\pm \alpha}}{\cosh \alpha}\right) = \pm \frac{1}{\cosh^2 \alpha}. \tag{70}$$

The time-like wave function (70) is that of a neutral particle, which is the free photon encountered in nature. Although the supposition, that the ground state photon has a charge, seems to be rather unphysical, the combination with its antiparticle forms a stable compound particle, which is the massive electron or positron.

The solution of the translational equation (11) represents the propagating carrier wave of the particle. For simplicity, we consider for the photon only the plane-wave solution

$$\vec{\psi}_{tp}(\vec{r}, t) = \vec{A}_{ph} = \vec{A}_{p0} e^{\pm i(\vec{k}\vec{r} - \omega t)}, \quad \vec{k}^2 = k^2 = \omega^2/c^2. \tag{71}$$

The three-dimensional polar vector $\vec{A}_{p0}$ determines the polarization of the photon wave. The vector $\vec{\psi}_{tr}(\vec{r}, t) = \vec{A}_{ph}$ corresponds to the classical magnetic vector potential and satisfies the Coulomb gauge. Present quantum mechanical theory considers the photon as a point-like particle. Therefore, it cannot account for the intrinsic angular momentum of the photon. The translational component (71) of the wave function takes only into account the polarization of the electromagnetic light wave. The total wave function of the free photon is the product of the angular wave function and the translational wave function (71). For the $l_s = 1$ state we obtain

$$\vec{\psi}_{ph} = \vec{\psi}_{tr}(\vec{r}, t) \psi_{p\alpha}^{(\pm)} = \vec{A}_p e^{i(\vec{k}\vec{r} - \omega t)} \frac{1}{\cosh^2 \alpha}. \tag{72}$$

In order to account for the polarization of the photon wave, the vector $\vec{A}_p$ must consist of at least two orthogonal components. The wave function (64) exhibits the wave-particle dualism of the photon. The direction of the angular momentum points in the direction of the wave-vector $\vec{k}$, which is perpendicular to the wave surface and determines its direction of propagation. We illustrate the dualistic behavior by considering that $\psi_{p\Omega}^{(\pm)}(\alpha, \theta, \phi)$ represents the wave function of a vortex or eddy. The eddy is riding on the propagating wave-surface like a surfer on the surf. The eddy represents the particle, which interacts with the detector, whereas the translational wave $\vec{\psi}_{tr}(\vec{r}, t)$ acts solely as the carrier wave determining the direct of motion of the eddy or photon,

The Standard Model of elementary particles assumes the existence of massive quarks, which have spin ½ and fractional charge $q = -\frac{1}{3}$ and $q = \frac{2}{3}$, respectively. In particular, the proton should consist of three quarks, two of them are "up" quarks ($q = 2/3$) and the third is the "down" quark [5]. Our model differs by supposing that massive quarks have integer charges and that each is composed of a massless quark and its antiparticle. Within the frame of our approach, at least two different massless quarks (quarkinos) $\nu = 1, 2$ exist with identical 4D helicity $s = \frac{1}{2}$, $j = \frac{1}{2}$, $\mu = \frac{1}{2}$. Their "*anti-quarkinos*" have opposit helicity $s = -\frac{1}{2}$, $j = -\frac{1}{2}$, $\mu = -\frac{1}{2}$. By considering these helicities, the wave functions of the quarkinos and those of their anti-quarkinos are

$$\Psi_{qv}^{(\pm)} = \psi_{Rv}(R)\Psi_{q\Omega}^{(\pm)}, \quad v = 1, 2 \qquad \Psi_{q\Omega}^{(\pm)} = e^{\pm\alpha/2}e^{\pm i\theta/2}e^{\pm i\phi/2}, \tag{73}$$

The wave functions $\psi_{R1}$ and $\psi_{R2}$ are the linearly independent solutions of the radial equation (41). For the quantum number $\mu = 1/2$ we obtain

$$\psi_{R1}(R) = \frac{C_1}{R^{1/2}} = \frac{\sqrt{3}}{R^{1/2}R_q^{3/2}}, \qquad \psi_{R2}(R) = \frac{C_2}{R^{3/2}} = \frac{1}{R^{3/2}R_q^{1/2}}. \tag{74}$$

The radius $R_q$ defines a characteristic elementary length, which determines the strength of the coupling between the quarkino and its antiquarkino. We have chosen the unknown factors $C_1$ and $C_2$ such that the square of each wave function (74) represents a 4D-density and $\int_0^{R_0} \psi_{Rv}^2 R^3 dR = 1$. We cannot normalize the radial wave functions (74) over the entire range $0 \leq R \leq \infty$ because the integrand $\psi_{Rv}^2 R^3$ diverges at the limit $R \to \infty$. Hence, it is not possible to assign the radial wave functions (74) to stable particles, which are at rest in an inertial system. Therefore, the quarkinos must be the quanta of some kind of radiation, although one has never observed them in any experiment [5, 12]. Because two radial solutions $\psi_{R1}(R)$ and $\psi_{R2}(R)$ exist for each of the spin-wave-functions $\Psi_{q\Omega}^{(\pm)}$, we suppose that the two particles are in a mixed radial state. We assign the symmetric radial wave function $\psi_R^{(+)} = (\psi_1 + \psi_2)/\sqrt{2}$ to the angular wave function $\Psi_{q\Omega}^{(+)}$ and the antisymmetric radial function $\psi_R^{(-)} = (\psi_1 - \psi_2)/\sqrt{2}$ to the angular wave function $\Psi_{q\Omega}^{(-)}$, giving

$$\Psi_{qr}^{(+)} = \left(\frac{\sqrt{3}}{R^{1/2}R_q^{3/2}} + \frac{1}{R^{3/2}R_q^{1/2}}\right)\frac{\Psi_{q\Omega}^{(+)}}{\sqrt{2}}, \qquad \Psi_{qr}^{(-)} = \left(\frac{\sqrt{3}}{R^{1/2}R_q^{3/2}} - \frac{1}{R^{3/2}R_q^{1/2}}\right)\frac{\Psi_{q\Omega}^{(-)}}{\sqrt{2}}. \tag{75}$$

Because the quarkino and the antiquarkino are independent particles, they cannot annihilate each other. Their mutual 4D density is

$$\varrho_q(R) = \Psi_{qr}^{(+)}\Psi_{qr}^{(-)} = \frac{1}{2}\left(\frac{3}{RR_q^3} - \frac{1}{R^3 R_2}\right). \tag{76}$$

We conceive each $\varrho_q$ as a 4D coupling density determining the strength of the interaction between the spin-1/2 quarkino and its antiparticle. Because the density (76) consists of a positive term and a negative term, it has the property of a hyper-spherical dipole. The 4D density is positive for the inner region $R > R_q$ and negative for the outer shell $R < R_q$.

## 6. Four-dimensional scalar potentials

We suppose that massless elementary particles are 4D vortices representing helical solitons. One characterizes vortices by their helicity, which is either left-handed or right-handed. Three-dimensional vortices observed in nature are tornados, moving smoke tori, and eddies riding on the surface of water waves. Vortices of opposite helicity attract each other whereas those of equal helicity repel each other. Accordingly, we suppose that particle and antiparticle have opposite 4D helicity. The source of the 4D potentials is the 4D particle-antiparticle coupling density

$$\varrho(R) = \psi_R^2 \Psi_\Omega \Psi_\Omega^\dagger = \psi_R^2 . \tag{77}$$

The inverse radius $1/R_c$ determines the strength of the coupling, the smaller the radius is the stronger is the coupling. Contrary to standard quantum theory, we do not interpret the density (77) as the probability density. Instead, we consider $\varrho(R)$ as the source of the four-dimensional hyper-symmetric scalar interaction potential $\Phi = \Phi(R)$, which accounts for the mass of the particle. In accordance with classical physics, we suppose that the 4D potential is the solution of the four-dimensional Poisson equation

$$\Delta_R^{(4)} \Phi = \frac{d^2\Phi}{dR^2} + \frac{3}{R}\frac{d\Phi}{dR} = \frac{1}{R^3}\frac{d}{dR}\left(\frac{d(R^3\Phi)}{dR}\right) = \varrho(R). \tag{78}$$

The density $\varrho(R)$ gives rise to a four-dimensional quantum well, which attracts the particles in the same way as the 4D gravitational potential of the sun bounds the earth according to general relativity.

Free photons and free anti-photons have opposite space-like helicity. As a result, free photons have the tendency to bunch up as demonstrated by the Hanbury-Brown-Twiss experiment [13]. To account for a significantly larger attraction the opposite time-like helicity of ground-state photon and anti-photon is the reason for their strong mutual attraction forming a stable particle, which is the electron and the positron, respectively. The 4D coupling density of two ground-state photons (69) with opposite 4D angular momentum (helicity) is

$$\varrho_p = \frac{4}{R_p^4}\Psi_{p\Omega}\Psi_{p\Omega}^\dagger = \frac{4}{R_p^4}. \tag{79}$$

By substituting $\varrho_p$ for $\varrho$ in (78) and integrating, we obtain

$$\Phi_{ph}(R) = \frac{R^2}{2R_p^4}. \tag{80}$$

We obtain the potential of the quarkino-antiquarkino compound particle by substituting the density (76) for $\varrho$ in the 4D Poisson equation (78), giving

$$\Phi_q(R) = \frac{1}{2}\left(\frac{R}{R_q^3} + \frac{1}{RR_q}\right). \tag{81}$$

The hyper-symmetric 4D potential (81) resembles the 3D potential of quantum chromodynamics (QCD), which diverges for short distances like the Coulomb potential and increases without limit at large distances to account for quark confinement [5]. However, the sign of the Coulomb-like term in (81) is not negative but positive. Therefore, the potential $\Phi_q(R)$ represents an entirely positive hyper-spherical potential well with infinitely high walls. As a result, any particle trapped within this potential cannot get out without some interaction.

## 7. Neutrinos

Neutrinos and antineutrinos are independent particles, which differ by their opposite spin. In particular, neutrinos are left-handed and have spin $s = 1/2$, whereas antineutrinos are right-handed having spin $s = -1/2$. The time-like (electric) component of the 4D angular momentum of electron neutrinos is zero. Therefore, they have entirely space-like (magnetic) property. As a result, electron neutrinos can only interact with the magnetic field via their magnetic moment $\vec{\mu}_\nu$. Unfortunately, we do not know its exact value, which must be very small owing to the very weak interaction of neutrinos with matter. Experiments with solar neutrinos have shown that $\mu_\nu \leq 2.8 \cdot 10^{-11} \mu_B$ [14]. As a result, the nature of the neutrino magneton $\mu_\nu$ must be completely different from that of the Bohr magneton $\mu_B$, which is proportional to $e/m_e$. Fortunately, within the frame of our approach, we avoid this discrepancy because the 4D angular momentum of the particle couples with the electromagnetic field. However, we do not know the coupling strength, which must be a factor $\leq 2.8 \cdot 10^{-11}$ smaller than that of the electron.

We postulate that free neutrinos and antineutrinos propagate with the velocity of light. As consequence, we must assume that free neutrinos are massless because otherwise their translational energy would be infinite. However, the neutrino gets a mass by the bonding of its magnetic moment with that of the nucleus. In this case, the translational motion of the center of gravity of the neutrino is at rest in the inertial system of the nucleus. According to our approach, neutrinos ($s = 1/2$) and antineutrinos ($s = -1/2$) can have different rotational eigenstates characterized by the polar angular quantum number $l_s$. The corresponding eigenfunctions are linearly independent solutions of the angular-momentum equation $\vec{M}^2 \psi_{l_s}^{(s)}(\alpha, \theta, \phi) = 0$. Free neutrinos are in the ground state $j = 0, l_s = 0, \alpha = 0$, In this case the time-like component of the angular tensor operator (38) is zero. The solutions of the resulting equation (58) for snin-1/2 particles in the case $j = 0$ are

$$\Psi_0^{(s)}(\theta, \phi) = \psi_0^{(s)}(\theta, \phi)\sqrt{\sin\theta} = e^{\pm i\theta/2} e^{is\phi}, \ s = \pm 1/2. \tag{82}$$

Because these solutions do not depend on the space-time angle $\alpha$, the angular momentum of the neutrino is of entirely magnetic (space-like) nature in the ground state, which the neutrino adopts if it is bound and at rest within a nucleus. Two polar solutions $e^{\pm i\theta/2}$ exist for each the neutrino and the antineutrino. The linear combinations $e^{is\phi}\sin\theta/2$ and $e^{is\phi}\cos\theta/2$ represent the components of the Dirac spinors for the free electron neutrino and antineutrino, respectively [15]. Because the rotational eigenfunctions of the higher states $l_s \geq 1$ do also depend on the time-like angle $\alpha$, they have space-like and time-like (electrical) properties:

$$\Psi_{l_s}^{(s)}(\alpha, \theta, \phi) = \psi_{l_s}^{(s)} \cosh\alpha \sqrt{\sin\theta} = \frac{C_{l_s-1}^{(1)}(\tanh\alpha)}{\cosh\alpha} e^{\pm i(l_s + \frac{1}{2})\theta} e^{is\phi}, \ l_s = 1, 2, \ldots . \tag{83}$$

The solutions (83) form a complete set of orthogonal eigenfunctions. Therefore, we may assign each of them to a different type of neutrino, for example the state $l_s = 0$ to the electron neutrino, $l_s = 1$ to the muon neutrino, $l_s = 2$ to the tau neutrino, and $l_s \geq 3$ to some other neutrinos, provided they exist. The eigenstates $l_s$ are rotational eigenstates, which we should not confuse with mass eigenstates. Because the eigenfunctions (83) depend on the angle $\alpha$ apart from the space-like angles $\theta$ and $\phi$, the neutrinos have also time-like properties for the states $l_s \geq 1$. However, the missing solar neutrinos contradict our proposition because the electron neutrino converts in flight to the muon neutrino and vice versa. A possible explanation of the experimental results may be the fact that the solutions (82) are not orthogonal to the eigenfunctions (83). Therefore, the solution (82) does not represent an independent eigenstate of the free neutrino. Instead, the electron neutrino may be in an oscillating mixed state, which must be a function of the Lorentz-invariant universal time. However, the oscillations can only start after the emission of the neutrino if an interaction exists between the states.

Instead of assigning a different type of neutrino to each quantum state $l_s$, we can interpret these states as precession states of the spin of the neutrino owing to their space-like and time-like properties. In the following, we restrict the spin precession (oscillation) to that between the states $l_s = 0$ and $l_s = 1$. The creation of the wave function $\Psi_1^{(s)}$ induced by the interaction with the state $l_s = 0$ results in a decrease of the initial space-like wave function $\Psi_{\nu 0}^{(s)}$ and vice versa. The frequency of the resulting oscillation or precession depends on the strength of the interaction, given by the interaction Hamilton operator $\widetilde{H}_{4i}^{(\pm)} = \pm k_\nu^2$, which acts as a creation or annihilation operator depending on the sign. The minus sign indicates an annihilation and the plus sign the creation of a state. Although alternating interactions do not introduce any coupling between the translational and the rotational motion, they induce an oscillation such that the entire wave function of the neutrino state is not static but stationary ($\partial/\partial\tau \neq 0$) oscillating between the mixed states $\Psi_{\tau 1}^{(s)}$ and $\Psi_{\tau 2}^{(s)}$. By considering that $\widetilde{H}_{4t}\psi = 0$ and $\widetilde{H}_{4r}\psi = 0$, we derive the alternating creation and annihilation of the states by the coupled equations

$$\frac{\partial \Psi_{\tau 1}^{(s)}}{\partial \tau} = -D_0 \widetilde{H}_{4i} \Psi_{\tau 2}^{(s)} = -D_0 k_\nu^2 \Psi_{\tau 2}^{(s)}, \quad \frac{\partial \Psi_{\tau 2}^{(s)}}{\partial \tau} = D_0 \widetilde{H}_{4i} \Psi_{\tau 1}^{(s)}. \tag{84}$$

The interaction Hamiltonian accounts for the transition from the initial 4D rotational state to another state and back again. Therefore, the mass of the neutrino stays zero and its velocity of light remains unaffected by the oscillations. We decouple the equations (84) by performing the second derivative, giving

$$\frac{\partial^2 \Psi_{\tau 1}^{(s)}}{\partial \tau^2} + \omega_\nu^2 \Psi_{\tau 1}^{(s)} = 0, \quad \frac{\partial^2 \Psi_{\tau 2}^{(s)}}{\partial \tau^2} + \omega_\nu^2 \Psi_{\tau 2}^{(s)} = 0, \quad \omega_\nu = D_0 k_\nu^2. \tag{85}$$

The mixed wave function $\Psi_{\tau 1}^{(s)}$ must satisfy the initial condition $\Psi_{\tau 1}^{(s)}(\tau_0) = \Psi_0^{(s)}$ and the wave function $\Psi_{\tau 2}^{(s)}$ the initial condition $\Psi_{\tau 2}^{(s)}(\tau_0) = -\Psi_{\tau 1}^{(s)\prime}(\tau_0)/\omega_\nu = -\Psi_{\nu 1}^{(s)}$. The solutions of (85), which fulfill these starting conditions, are the linear combinations

$$\Psi_{\tau 1}^{(s)}(\tau) = \Psi_{\nu 0}^{(s)} \cos \omega_\nu (\tau - \tau_0) - \Psi_{\nu 1}^{(s)} \sin \omega_\nu (\tau - \tau_0), \tag{86}$$

$$\Psi_{\tau 2}^{(s)}(\tau) = -\frac{1}{\omega_\nu} \frac{\partial \Psi_{\tau 1}^{(s)}}{\partial \tau} = \Psi_{\nu 1}^{(s)} \cos \omega_\nu (\tau - \tau_0) + \Psi_{\nu 0}^{(s)} \sin \omega_\nu (\tau - \tau_0). \tag{87}$$

The orthogonal wave functions (86, 87) of the mixed stationary states show that the neutrino will convert from the sate state $\Psi_{\nu 0}^{(s)}$ to the state $\Psi_{\nu 1}^{(s)}$ and then back again as it is the case for coupled pendula. So far, we have restricted our approach to oscillations between the ground state and the next higher state. However, we cannot exclude the possibility of oscillations between more than two states. The probability to detect the neutrinos in one of the rotational states $\Psi_{\nu 0}^{(s)}$ and $\Psi_{\nu 1}^{(s)}$ is 50%. This value is in reasonable agreement with the experimental value obtained by Fukuda et al. [16]. Their Super-Kamiokande detector counted about 45% of the solar neutrinos. We can explain the difference by assuming another interaction Hamilton operator $\widetilde{H}_{4i}^{(\pm)} = \pm k_\tau^2$ accounting for an oscillation between the neutrino states $l_s = 0$ and $l_s = 2$. We may exclude oscillations between the muon and the tau neutrino because their eigenfunctions are orthogonal to each other. The oscillatory eigenstates (86) are equivalents of the neutrino flavor eigenstates in QCD whose structure is largely unknown.

## 8. Massive two-component systems

The Standard Model of elementary particles assumes that the quarks are initially massless and that they obtain their mass by coupling with the Higgs particle [6, 7]. We propose an alternative model, which postulates that stable massive elementary particles result from the mutual coupling of mass-less particles, such as photons and time-like pseudo-quarks. The masses of the compound system are eigenvalues of its four-dimensional rotational energy. The simplest massive particle consists of two massless particles. In the following, we restrict our investigations to two-component systems because we can solve the wave equation of such systems for the relevant interaction potentials analytically. Promising candidates representing two-component systems are the electron, the proton, and their antiparticles. We suppose that the Hamiltonian $\widetilde{H}_{4i}$, which accounts for interaction between the particles of two-body system, exhibits the highest possible four-dimensional symmetry. This is the case if the modified potential is hyper-spherically symmetric depending only on the 4D-distance $R = |\vec{R}_1 - \vec{R}_2|/2$ of each particle from their common "center of gravity":

$$\widetilde{H}_{4i} = \Phi(\vec{R}_1, \vec{R}_2) = \Phi(|\vec{R}_1 - \vec{R}_2|) = \Phi(R). \tag{88}$$

Because the potentials (80) and (81) are of the hyper-symmetric form (80), they are the most promising potentials for the constituents of massive elementary particles. The hyper-symmetry of the 4D potentials (80) and (81) enables one to separate the translational motion of the center of gravity from the rotational motion of the two-body system in the 4D-space by introducing the center of gravity coordinates $x_\nu = (x_{1\nu} + x_{2\nu})/2$ and the relative coordinates $x_{r\nu} = (x_{1\nu} - x_{2\nu})/2, \nu = 1,2,3,4$. We describe rotations in the 4D-space by means of the hyper-spherical coordinates $R, \alpha, \theta, \phi$. The coordinates $x_\nu$ define the position of the center of the compound particle whereas the relative coordinates $x_{r\nu}$ determine the distance of each constituent particle from the center of gravity. The relations existing between the relative Cartesian coordinates and the hyper-spherical coordinates are

$$x_{r4} = \frac{1}{2}(x_{14} - x_{24}) = iR \sinh \alpha,$$

$$x_{r3} = \frac{1}{2}(x_{13} - x_{32} = R \cosh \alpha \cos \theta,$$

$$x_{r2} = \frac{1}{2}(x_{12} - x_{22}) = R \cosh \alpha \sin \theta \sin \phi,$$

$$x_{r1} = \frac{1}{2}(x_{11} - x_{21}) = R \cosh \alpha \sin \theta \cos \phi. \tag{89}$$

The hyper-symmetric coordinates $x_{rv}$ can adopt positive and negative values because the property of the compound particle cannot depend on the numbering of its constituents.

By introducing the hyper-spherical coordinates (89) and assuming, the wave equation of the static ($\partial/\partial \tau = 0$) two-particle system is

$$-\widetilde{H}_1 \psi - \widetilde{H}_2 \psi - \widetilde{H}_i \psi = -\frac{1}{2}\widetilde{H}_t \psi - \frac{1}{2}\widetilde{H}_r \psi - \widetilde{H}_{4i} \psi =$$

$$\frac{1}{2}\left(\Delta_t^{(4)} + \Delta_r^{(4)}\right)\psi(x_v, R, \alpha, \theta, \phi) - \Phi(R)\psi(x_v, R, \alpha, \theta, \phi) = 0. \tag{90}$$

The hyper-symmetric scalar interaction potential $\Phi(R)$ only affects the rotational motion. Therefore, we can separate the motion of the center of gravity from the four-dimensional rotation by factorizing the wave function in the form

$$\psi(x_v, R, \alpha, \theta, \phi) = \psi_t(x_v) \cdot \psi_r(R, \alpha, \theta, \phi). \tag{91}$$

The wave function $\psi_r(R, \alpha, \theta, \phi)$ accounts for the rotation of the 4D compound particle. The substitution of (86) for $\psi(x_v, R, \alpha, \theta, \phi)$ in (90) enables us to separate the translational motion from the rotational motion by employing the Bernoulli separation procedure, giving

$$\widetilde{H}_{4r}\psi + \widetilde{H}_{4i}\psi = -\Delta_r^{(4)}\psi_r + 2\Phi\psi_r = -\Lambda\psi_r = \widetilde{E}_r^{(4)}\psi_r, \quad \psi_r = \psi_r(R, \alpha, \theta, \phi), \tag{92}$$

$$\widetilde{H}_{4t}\psi = -\Delta_t^{(4)}\psi_t = \Lambda\psi_t = \widetilde{E}_t^{(4)}\psi_t \rightarrow \Delta_t^{(4)}\psi_t - \Lambda\psi_t = 0, \quad \psi_t = \psi_t(x_v). \tag{93}$$

Equation (87) accounts for the four-dimensional rotation of the system and (93) for the propagation of its center of gravity. In the case $\Lambda = \widetilde{E}_t^{(4)} = -k_C^2$, $k_C = m_e c/\hbar$, equation (93) represents the Klein-Gordon equation. In this case, the eigenvalue of the 4D rotational energy $\tilde{E}_r^{(4)} = -\Lambda = -\tilde{E}_t^{(4)} = k_C^2$ has space-like property, which implies that the 4D rotational system is closed. The separation constant is an eigenvalue of the 4D rotational equation (93) determining the mass of the compound particle. The absolute values of the rotational 4D energy and the translational 4D energy are equal. The total 4D energy of the particle $E^{(4)} = \tilde{E}_r^{(4)} - \tilde{E}_t^{(4)} = 2\,\tilde{E}_r^{(4)}$ is positive and nonzero, as it is the case for the electric and magnetic energy of electromagnetic waves. The equivalence strongly supports the physical relevance of the new model. Moreover, because the angular operators have the same properties as the components of the electromagnetic field tensor, we suppose that the rotational energy represents the finite field energy of the particle.

The Klein-Gordon equation accounts for relativistic effects but cannot accommodate the spin-1/2 nature of the electron [16]. Our model describes the intrinsic rotation (spin) of the compound particle by the wave function $\psi(R, \alpha, \theta, \phi)$, which accounts for the four-dimensional hyper-spin of the particle. Therefore, we suppose that our approach is a promising alternative to the Dirac formalism [14] and to the Standard Model for describing the properties of elementary particles. Dirac considered the electron as a point-like particle with an inherent spin. To account for the spin, the wave function of the electron is a four-component spinor wave function. We propose an alternative formulation by assuming a single scalar wave function, which depends on the eight variables $x_\nu, R, \alpha, \theta, \phi$. Moreover, we consider the electron as a 4D rotating top with a finite 4D-radius. This assumption is in agreement with classical mechanics, according to which a rotating system must have a finite radius in order that the rotational energy does not diverge, as it is the case for tornados and hurricanes. Equations (92) and (93) show that the 4D translational energy $\tilde{E}_t^{(4)}$ and the rotational energy $\tilde{E}_r^{(4)} = \Lambda$ of the compound particle are equal. Hence, the total 4D energy of the particle is twice its 4D translational energy. Because the 4D rotational energy is time-like, we suppose that the 4D rotation is the source of the electromagnetic field of the particle.

To find the eigenfunctions and eigenvalues for the potentials (80) and (81), we must look for separable solutions of the rotational 4D Hamilton operator for spin-1/2 particles. Separable wave functions in hyper-spherical coordinates have the form

$$\psi_r = \psi_R(R)\psi_\alpha(\alpha)\psi_\theta(\theta)\psi_\phi(\phi) = \psi_R(R)\frac{\Psi_\alpha(\alpha)}{\cosh\alpha}\frac{\Psi_\theta(\theta)}{\sqrt{\sin\theta}}e^{is\phi}, \quad s = \pm 1/2. \qquad (94)$$

By introducing the modified wave functions $\psi_R(R), \Psi_\alpha(\alpha)$ and $\Psi_\theta(\theta)$, we can separate equation (92), giving equation (48) for $\Psi_\alpha(\alpha)$, (49) for $\Psi_\theta(\theta)e^{is\phi}$ and for $\psi_R(R)$ the radial equation

$$\frac{d^2\psi_R(R)}{dR^2} + \frac{3}{R}\frac{d\psi_R(R)}{dR} + \{\Lambda - 2\Phi\}\psi_R(R) + \frac{3}{4R^2}\psi_R(R) = 0. \tag{95}$$

The normalization condition factorizes in a product of single integrals in the form

$$\int \psi_r \psi_r^\dagger dV_r^{(4)} = \int_0^\infty \int_{-\infty}^\infty \int_0^\pi \int_0^{2\pi} \int \psi_r \psi_r^\dagger R^3 \cosh^2\alpha \, dRd\Omega^{(4)} =$$

$$\int_0^\infty \int_{-\infty}^\infty \int_0^\pi [\psi_R(R)]^2 \, \Psi_\alpha \Psi_\alpha^\dagger \Psi_\theta \Psi_\theta^* R^3 dR \frac{d\alpha}{\cosh\alpha} d\theta = 1. \tag{96}$$

### 9. Electron and Positron

The electron and the positron are stable spin-1/2 particles, which differ by the sign of their charge. They annihilate each other giving two photons with spin $\pm 1$. The electromagnetic vector potential represents the wave function of the photon. We obtain the wave functions of the electron and the positron by assuming that each of them consists of two entangled photons with opposite spin $\pm 1$ bound in a common singlet state. Accordigly, the annihilation of an electron positron pair produces entangled photons keeping their initial state. This assumption explains the close connection between photons and the electron and positron.

The mutual vortex density of each photon pair produces the 4D-potential (80). By substituting (38) for $\Delta_r^{(4)}$ and $\Phi_{ph}(R) = R^2/8R_p^4$ for $\Phi(R)$ in (92), we obtain the rotational wave equation

$$(\Lambda - \widetilde{H}_{4r})\psi_r = \frac{\partial^2 \psi_r}{\partial R^2} + \frac{3}{R}\frac{\partial \psi_r}{\partial R} - \frac{R^2}{4R_p^4}\psi_r + \Lambda\psi_r$$

$$-\frac{1}{R^2}\left(\frac{\partial^2 \psi_r}{\partial \alpha^2} + 2\tanh\alpha \frac{\partial \psi_r}{\partial \alpha} - \frac{1}{\cosh^2\alpha}\left\{\frac{1}{\sin\theta}\frac{\partial}{\partial \theta}\left(\sin\theta \frac{\partial \psi_r}{\partial \theta}\right) + \frac{1}{\sin^2\theta}\frac{\partial^2 \psi_r}{\partial \phi^2}\right\}\right) = 0. \tag{97}$$

This equation has solutions, which factorize in the form (94). In the special case $\mu = \pm 1/2$, the Bernoulli separation procedure gives the equations

$$\frac{d^2 \Psi_R}{dR^2} - \left[\frac{R^2}{4R_p^4} - \Lambda\right]\Psi_R = 0, \qquad \Psi_R = \psi_R R^{3/2}, \tag{98}$$

$$\frac{\partial^2 \Psi_\alpha}{\partial \alpha^2} - \frac{1}{4}\Psi_\alpha - \frac{l_s(l_s+1)}{\cosh^2\alpha}\Psi_\alpha = 0, \qquad l_s = 0, 1, 2, \ldots, \tag{99}$$

$$\frac{d^2\Psi_\theta}{d^2\theta} + j_s{}^2\Psi_\theta = 0, \qquad j_s = l_s + \frac{1}{2} = \frac{1}{2}, \frac{3}{2}, \dots . \tag{100}$$

The radial equation (98) is that of the harmonic 4D oscillator, which has nomalizable eigenfunctions $\Psi_{Rn}, n = 0, 1, 2, ..$, for a set of eigenvalues

$$\Lambda = \Lambda_n = \Lambda_n = \frac{2n+1}{R_0^2} = (2n+1)k_C^2, \quad n = 0, 1, 2, \dots . \tag{101}$$

The associated eigenfunctions are

$$\psi_{Rn}(u) = \frac{\Psi_R}{u^{3/2}} = C_{Rn} \frac{e^{-\frac{u^2}{2}}}{u^{\frac{3}{2}}} H_n(u), \quad u = R/R_0, \quad R_0 = 2R_p = \frac{1}{k_C} = \alpha_S r_0. \tag{102}$$

The polynomials $H_n(u)$ are the Hermit polynomials of order $n$; $\alpha_S$ and $r_0$ are the Sommerfeld fine-structure constant and the radius of the hydrogen atom, respectively. We have equalized the elementary diameter $R_0$ with the inverse of the Compton wave number $k_C = mc/\hbar$ in order to obtain the Klein-Gordon equation for the translational motion of the electron. We assign the radial wave functions (102) to both the electron and to the positron. Moreover, we interpret the excited states $n > 0$ as those of the electron or positron plus $n$ electron-positron pairs. For the ground state $n = 0$, we obtain the mean square 4D radius as

$$\langle R^2 \rangle = \frac{1}{2}R_0^2 = \frac{1}{2k_C^2}. \tag{103}$$

The total 4D energy is

$$E_n^{(4)} = \left(\widetilde{E}_{rn}^{(4)} - \widetilde{E}_{tn}^{(4)}\right)\hbar^2 c^2 = 2\widetilde{E}_{rn}^{(4)}\hbar^2 c^2 = 2(2n+1)\hbar^2 k_C^2 c^2 = 2m_0^2 c^4 + n(2m_0 c^2)^2 \tag{104}$$

The Lorentz invariant 4D energy of the particle is constant regardless of the chosen inertial system and should not be confused with the three-dimensional rest energy $E_0 = m_0 c^2$. We can interpret the result (104) as the 4D energy of the electron plus the 4D energy of $n$ electron positron pairs. In the case of a single ($n = 0$) electron or positron, half of the 4D energy $2m_0^2c^4$ results from the time-like 4D translational motion the other half from the space-like 4D rotation. Our approach gives a distinct finite value for the field energy of the particle and thus avoids the divergence problem encountered in classical electrodynamics. The similarity of our approach with classical electrodynamics becomes obvious by considering that the energy density $\vec{B}^2 + \vec{E}^2/c^2 > 0$ of electromagnetic waves is nonzero, whereas the Lorentz invariant Lagrange density $\sum_{\mu<\nu} F_{\mu\nu}^2 = \vec{B}^2 - \vec{E}^2/c^2 = 0$. In order to obtain a finite field energy for the electron, one introduces in classical electrodynamics the electron radius $r_e = \alpha_S/k_C$ obtained by equalizing the Coulomb energy with the rest energy of the electron.

Equation (99) has relevant solutions, which describe time-like rotations:

$$\psi_{\alpha l_s}^{(\pm)} = \psi_{l_s}^{(\pm)}(\alpha) = \Psi_{l_s}^{(\pm)}(\alpha)/\cosh\alpha = \Psi_{\alpha l_s}^{(\pm)}/\cosh\alpha =$$

$$C_{l_s} \cosh^{l_s}\alpha \left(\frac{1}{\cosh\alpha}\frac{d}{d\alpha}\right)^{l_s} \frac{e^{\pm\frac{\alpha}{2}}}{\cosh\alpha} = e^{\pm\alpha/2} D_{l_s}^{(\pm 1/2)}(\tanh\alpha), \quad l_s = 0, 1, 2 \dots . \quad (105)$$

The normalization constant $C_{l_s}$ is the same for $\psi_{\alpha l_s}^{(+)}$ and $\psi_{\alpha l_s}^{(-)}$. The polynomials $D_{l_s}^{(\pm 1/2)}(\tanh\alpha)$ do not form a set of orthogonal eigenfunctions. Therefore, states with a fixed quantum number $l_s > 0$ are not stable apart from the ground state $l_s = 0$. The normalized wave functions (105) for the ground state $l_s = 0$ and for the state $l_s = 1$ are

$$\psi_{\alpha 0}^{(\pm)} = \frac{1}{\sqrt{\pi}}\frac{e^{\pm\frac{\alpha}{2}}}{\cosh\alpha}, \qquad \psi_{\alpha 1}^{(\pm)} = \frac{2}{\sqrt{\pi}}\frac{e^{\pm\frac{\alpha}{2}}}{\cosh\alpha}\left(\pm\frac{1}{2} - \tanh\alpha\right). \quad (106)$$

The wave functions $\psi_{l_s}^{(\pm)}(\alpha)$ produced by successive application of the rotation operator $(1/\cosh\alpha)\partial/\partial\alpha$ are eigenfunctions of the time-like spin $s_t = \pm 1/2$. The sign of the time-like spin defines two different elementary particles, as it is the case for the neutrinos where the neutrino has spin eigenvalue $s = 1/2$ and the antineutrino $s = -1/2$. Therefore, we postulate that the minus sign accounts for the electron and the plus sign for its antiparticle, the positron. By differentiating (105), we obtain the relation

$$\psi_{\alpha,l_s+1}^{(\pm)} = C_{l_s+1}\left(\frac{\partial\psi_{\alpha,l_s}^{(\pm)}}{\partial\alpha} - (l_s+1)\tanh\alpha\,\psi_{\alpha,l_s}^{(\pm)}\right). \quad (107)$$

The operator $\frac{\partial}{\partial \alpha} - (l_s + 1) \tanh \alpha$ is a creation operator, which rises the time-like angular state $\psi_{\alpha,l_s}^{(\pm)}$ of the particle to the next higher state $\psi_{\alpha,l_s+1}^{(\pm)}$.

Equation (100) has the linearly independent solutions $\Psi_\theta = \cos j_s \theta$ and $\Psi_\theta = \sin j_s \theta$. By employing these solutions, we obtain the normalized angular eigenfunctions

$$\psi_{j_s}^{(\uparrow)}(\theta) = \sqrt{\frac{2}{\pi}} \frac{\cos j_s \theta}{\sqrt{\sin \theta}}, \quad \psi_{j_s}^{\downarrow}(\theta) = \sqrt{\frac{2}{\pi}} \frac{\sin j_s \theta}{\sqrt{\sin \theta}}, \quad j_s = \frac{1}{2}, \frac{3}{2}, \dots . \tag{108}$$

We assign the wave function $\psi_{j_s}^{(\uparrow)}(\theta)$ to the "spin-up" state and $\psi_{j_s}^{(\downarrow)}$ to the "spin-down" state because the first function maximizes at $\theta = 0$, the other at $\theta = \pi/2$. Free electrons and free positrons are in the ground state $n_s = 0$, $l_s = 0$, $j_s = 1/2$. We readily obtain the normalized rotational wave functions $\psi_{re}^{(\uparrow)}$ and $\psi_{re}^{(\downarrow)}$ of the free ($j_s = 1/2$) electron from (10), (94), (101), (106), and (108) as

$$\psi_{re}^{(\uparrow)} = \frac{\sqrt{k_C}}{\pi \sqrt[4]{4\pi}} \frac{e^{-\frac{R^2 k_C^2}{2}}}{R^{\frac{3}{2}}} \frac{e^{-\frac{\alpha}{2}}}{\cosh \alpha} \sqrt{\cot \theta/2}\, e^{i\phi/2},$$

$$\psi_{re}^{(\downarrow)} = \frac{\sqrt{k_C}}{\pi \sqrt[4]{4\pi}} \frac{e^{-\frac{R^2 k_C^2}{2}}}{R^{\frac{3}{2}}} \frac{e^{-\frac{\alpha}{2}}}{\cosh \alpha} \sqrt{\tan \theta/2}\, e^{-i\phi/2}. \tag{109}$$

We obtain the corresponding wave functions of the free positron by substituting $\alpha$ for $-\alpha$ in (109). As a result, the electron has negative time-like helicity, whereas the positron has positive time-like helicity. The spin-up state has positive spin $s = 1/2$, whereas the spin-down state has negative spin $s = -1/2$. The positron wave function $\psi_{rp}$ is the adjoint wave function $\psi_{re}^\dagger$ of the electron and vice versa.

## 10. Propagation in free space

The translational equation (85) determines the propagation of the compound particle in free space. For the electron and the positron, the equation represents the Klein-Gordon equation. This identity readily follows from the relation $\Lambda = k_C^2$, where $k_C = m_e c/\hbar$ is the Compton wave number of the electron and $m_e$ the rest mass. In free space, the translational motion is stationary with respect to the laboratory time $t = x_4/c$. By writing the wave function $\psi_t(\vec{r}, t)$ as a superposition of plane waves, the general solution of equation (88) is

$$\psi_t(\vec{r},t) = \iiint a(\vec{k})e^{i(\vec{k}\vec{r}+k_4x_4)}d^3k, \tag{110}$$

$$\vec{k}\vec{r} = \sum_{\nu=1}^{3} k_\nu x_{t\nu}, \quad k_4 x_4 = -\omega t. \tag{111}$$

The Fourier amplitude $a(\vec{k})$ determines the three-dimensional spatial structure of the wave. The time-like component $k_4 = i\omega/c$ of the four-dimensional wave vector depends linearly on the frequency of the wave. By substituting (110) for $\psi_t$ in (90), we obtain for the free particle the relation

$$\sum_{\nu=1}^{4} k_\nu^2 = \vec{k}^2 + k_4^2 = k^2 - \frac{\omega^2}{c^2} = -\Lambda. \tag{112}$$

By setting $\Lambda = 0$ in (112), it readily follows that massless particles always propagate with the velocity of light, whereas massive particles can propagate with velocity smaller than the velocity of light. In this case, an inertial system exists in which the particle is at rest ($k^2 = 0$) regardless how small its mass is.

The group velocity of the wave package (110) is the velocity $v$ of the particle in the three-dimensional space. By employing the relation (112) we obtain

$$\vec{v} = \vec{v}_g = \vec{\nabla}_k \omega = c\frac{\vec{k}}{\sqrt{\Lambda+k^2}}. \tag{113}$$

The translational motion is time-like in the case $\Lambda = (m_0 c/\hbar)^2 > 0$. $(m_0 c/\hbar)^2$, we rewrite (113) in the form

$$\hbar^2 \vec{k}^2 = \frac{\vec{v}^2}{c^2}\frac{\hbar^2 \Lambda}{1-v^2/c^2} = \frac{m_0^2 \vec{v}^2}{1-v^2/c^2} > 0. \tag{114}$$

From (114) and (112) we readily obtain the familiar relations for the components of the relativistic 4D kinetic momentum of the particle

$$\vec{p}_k = \hbar \vec{k} = m\vec{v} = \frac{m_0 c\vec{\beta}}{\sqrt{1-\beta^2}}, \quad p_4 = \hbar k_4 = i\frac{E}{c} = \frac{im_0 c}{\sqrt{1-\beta^2}}, \quad \beta = \frac{v}{c}, \quad m = \frac{m_0}{\sqrt{1-\beta^2}}. \tag{115}$$

The relation for the time-like component $p_4$ of the 4D translational momentum represents Einstein's formula for the 3D energy of a particle.

We cannot exclude the possibility that we obtain negative eigenvalues $\Lambda = -\lfloor \Lambda \rfloor$ for some other potentials. In this case, we obtain the peculiar result

$$\vec{p}_k = \hbar \vec{k} = m\vec{v} = \frac{m_0 c \vec{\beta}}{\sqrt{\beta^2 - 1}}, \quad p_4 = i\frac{E}{c} = \frac{im_0 c}{\sqrt{\beta^2 - 1}}, \quad \beta = \frac{v}{c} > 1, \quad m = \frac{m_0}{\sqrt{\beta^2 - 1}}. \tag{116}$$

In order that the kinetic momentum is real, the velocity $v = \beta c$ must be larger than the velocity of light. One calls such hypothetical particles "tachyons". Tachyons must propagate in the space-like region of the Minkowski space if they exist. So far, one has assumed that this region does not have any physical relevance because the known particles propagate in the time-like region. The light cone separates both regions from each other. Although we cannot see tachyons, they may account for dark matter, which interacts with the particles in the time-like region via gravitation.

## 11. Electron propagating in the electromagnetic field

In accordance with the standard Dirac theory, we assume an external non-quantized electromagnetic field. This approximation is valid in the steady case, where the large number of photons effectively smooth out the lumpiness of the field. In the presence of an external electromagnetic field ($\vec{A}, A_4 = iV/c$), we must replace the components of the kinetic momentum by the components $p_{c\mu} = p_{k\mu} + eA_\mu = \frac{\hbar \partial}{i \partial x_\mu} + eA_\mu$ of the four-dimensional canonical-momentum operator. We propose in close analogy that the equivalent replacement exists for the components (23) of the angular momentum operator. By supposing that the electromagnetic field strength induces an additional angular moment $\overleftrightarrow{M}^{(i)}$ with components $M_{\mu\nu}^{(i)} = eR^2 F_{\mu\nu}$, we obtain for the components $M_{\mu\nu}^{(c)}$ of the *canonical* angular momentum operator the relation

$$M_{\mu\nu}^{(c)} = M_{\mu\nu} + M_{\mu\nu}^{(i)} = i\hbar \Omega_{\mu\nu} + eR^2 F_{\mu\nu}. \tag{118}$$

The canonical angular momentum is gauge invariant, whereas the canonical kinetic momentum is not.

In the presence of an external electromagnetic field, the 4D translational Hamilton operator (11) is

$$\widetilde{H}_{4t} = \sum_{\mu=1}^{4}\left(\frac{\partial}{i\partial x_\mu} - \frac{e}{\hbar}A_\mu\right)^2 = -\Delta + \frac{1}{c^2}\frac{\partial^2}{\partial t^2} + 2\frac{e}{\hbar}\left(i\vec{A}\vec{\nabla} + \frac{A_4}{c}\frac{\partial}{\partial t}\right) + \frac{e^2}{\hbar^2}\vec{A}^2 + \frac{e^2}{\hbar^2}A_4^{\;2}. \quad (119)$$

Its time-like component $A_4 = -iV_e/c$ is proportional to the electric potential $V_e$. Apart from the interaction of the translational momentum with the vector potential, the intrinsic 4D angular momentum interacts with the electromagnetic field strengths via the rotation Hamiltonian

$$\widetilde{H}_r = -\frac{1}{R^2}\sum_{\mu<\nu}\left(i\Omega_{\mu\nu} + \frac{e}{\hbar}R^2 F_{\mu\nu}\right)^2 - \frac{\partial^2}{\partial R^2} - \frac{3}{R}\frac{\partial}{\partial R} - \Phi(R). \quad (120)$$

The part $\widetilde{H}_{ir}$ of the rotation Hamiltonian (120), which accounts for the interaction with the external electromagnetic field, is

$$\widetilde{H}_{ir} = -\frac{e}{\hbar}\sum_{\mu,\nu}F_{\mu\nu}\Omega_{\mu\nu} - \frac{e^2 R^2}{2\hbar^2}\sum_{\mu,\nu}F_{\mu\nu}^2 = 2i\frac{e}{\hbar}\left(\frac{1}{c}\vec{E}\cdot\vec{\Omega}_t - \vec{B}\cdot\vec{\Omega}_s\right) + \frac{e^2 R^2}{\hbar^2}\left(\frac{\vec{E}^2}{c^2} - \vec{B}^2\right). \quad (121)$$

The first term is the equivalent of the Dirac interaction Hamiltonian, which we obtain by iteration of the Dirac equation. This term accounts for the coupling of the components $F_{\mu\nu}$ of the electromagnetic field tensor with those of the angular momentum tensor. We readily show this equivalence by substituting the Dirac matrices $\sigma_{\mu\nu} = -i\gamma_\mu\gamma_\nu$ for the components $\Omega_{\mu\nu}$ in (125). The second term describes the scalar coupling of the square of the external electromagnetic field tensor with the 4D *moment of inertia* of the electron. The Dirac formalism does not know this term because it assumes a point-like particle ($R = 0$). The electric operator $\vec{E}\cdot\vec{\Omega}_t$ accounts for the energy of the spin-orbit coupling resulting from the interaction of the electric moment (time-like component of the 4D-spin) of the particle with the electric field. The interaction term of the translational Hamiltonian (119) is not entirely defined quantity because the electromagnetic vector potential is not gauge invariant. On the other hand, the rotational interaction Hamiltonian (121) is gauge invariant. Hence, it does not depend on the chosen inertial system. Although the operator (121) is non-Hermitian adding an imaginary term to the radial wave equation, it has realistic physical consequences as demonstrated in the next chapter.

The operator $i\vec{B}\cdot\vec{\Omega}_s$ acting on the spin eigenfunction $\psi_\phi$ gives a real value for the energy resulting from the interaction of the magnetic moment of the electron with the magnetic field. The z-component of the space-like operator $\vec{\Omega}_s$ is parallel to the direction of the magnetic field. In the case of an external homogeneous magnetic field $\vec{B} = B\vec{e}_z$ we obtain

$$\widetilde{H}_{im}\psi_\phi = -i\frac{2e}{\hbar}\vec{B}\cdot\vec{\Omega}_s\psi_\phi - \frac{e^2R^2}{\hbar^2}\vec{B}^2\psi_\phi =$$

$$-i\frac{2e}{\hbar}B\frac{\partial\psi_\phi}{\partial\phi} - \frac{e^2R^2}{\hbar^2}B^2\psi_\phi \approx \left(\frac{2e}{\hbar}Bs - \frac{e^2}{\hbar^2 k_C^2}B^2\right)\psi_\phi, \quad s = \pm\frac{1}{2}. \tag{122}$$

The sign of the linear term differs for the "up" (parallel) and "down" (antiparallel) spin. The quadratic term resulting from the introduction of the canonical angular momentum is negligibly small in the case of macroscopic magnetic fields.

## 12. The hydrogen atom revisited

So far, we have only considered the motion of elementary particles in the field free space. Charged particles propagating in the electromagnetic field radiate, whereas they do not radiate if they are in bound states within atoms. The Schrödinger equation explains this fact and provides an analytical solution for the energy states of the hydrogen atom. However, the energy states are highly degenerated, giving the same energy for different states. The Dirac equation reduces the degeneracy to a large extend but does not predict the Lamb shift [18]. Moreover, the probability density of the s-states of the hydrogen atom has a singularity at the origin. This divergence is unrealistic. On the other hand, the new model gives a probability density, which vanishes at the origin.

In the absence of the magnetic field ($\vec{B} = 0$), only the time-like component $A_4$ of the 4D vector potential is nonzero. For a point charge with atomic number $Z$, the 4D vector potential and the electric field strength $\vec{E}$ are

$$\vec{A} = 0, \quad A_4 = \frac{ie}{4\pi\varepsilon_0 c}\frac{Z}{r}, \quad \frac{i}{c}\vec{E} = -\vec{\nabla}A_4 = \frac{iZe}{4\pi\varepsilon_0 c}\frac{\vec{e}_r}{r^2}, \quad \frac{\vec{E}^2}{c^2} = \frac{\hbar^2}{e^2}\frac{Z^2\alpha_S^2}{r^4}. \tag{123}$$

By substituting (32) for the time-like operator $\vec{\Omega}_t$ and (123) for the electric field strength $\vec{E}$ in the rotational interaction operator (121), we obtain

$$\widetilde{H}_{ir} = 2i\frac{Z\alpha_S}{r^2}\frac{\partial}{\partial\alpha} + R^2\frac{Z^2\alpha_S^2}{r^4}. \tag{124}$$

Here $\alpha_S = e^2/4\pi\varepsilon_0\hbar c$ is the Sommerfeld fine-structure-constant. The first term is associated with the transition of the time-like angular momentum from one state to the next higher state, as follows from (59). The imaginary interaction term accounts for the energy of the photon, which couples the electron with the electric field. As a result, the wave function of the bound electron must depend on the universal time $\tau$ because the operator $\partial/\partial\tau$ of the wave function (5) accounts for the creation or annihilation of bound massless particles. Massless particles travel with the velocity of light. In this case, no inertial system exists in which the particles are at rest. In order to be at rest in the atom, they mass resulting from the bonding.

In the stationary case, the wave function factorizes in the form

$$\psi(\vec{r}, R, \alpha, \tau) = \psi(r, \vartheta, \varphi, R, \alpha)e^{i\omega_\tau \tau}, \quad \omega_\tau = D_0\widetilde{E}_r^{(4)}. \tag{125}$$

We describe the quantum state of the electron within the three-dimensional Coulomb field by the spherical coordinates $r, \vartheta, \varphi$. These coordinates differ from the hyper-spherical coordinates $R, \alpha, \theta, \phi$, which describe the intrinsic structure of the electron. We indicate the three-dimensional *orbital* angular momentum by the standard quantum numbers $l$ and $m$. The interaction Hamiltonian (124) couples the electron with the electric fired of the nucleus in such a way that we cannot decouple the 4D Hamiltonian in a term depending exclusively on the center-of-mass (translational) coordinates and a term depending solely on the intrinsic coordinates. Fortunately, we can decouple the spatial rotations. As a result, the eigenfunctions factorize in the form

$$\psi = \frac{\Psi(r,R,\alpha)}{R^{3/2}\cosh\alpha} Y_{l_o}^m(\vartheta, \varphi) \frac{e^{-ij_s\theta}}{\sqrt{\sin\theta}} e^{is\phi} e^{-i\omega t} e^{i\omega_\tau \tau}, \quad \omega = kc.\ j_s = l_s + 1/2. \tag{126}$$

We obtain the equation for the wave function $\Psi(r, R, \alpha)$ by inserting (126) for $\psi$ into the equation (4), giving

$$\frac{1}{D_0}\frac{\partial\psi}{c\partial\tau} = i\widetilde{E}_\tau^{(4)}\Psi = -(\widetilde{H}_{4t} + \widetilde{H}_{4r} + \widetilde{H}_{4i})\Psi, \tag{127}$$

$$-\widetilde{H}_{4t}\Psi = \frac{\partial^2\Psi}{\partial r^2} + \frac{2}{r}\frac{\partial\Psi}{\partial r} + k^2\Psi - \frac{l(l+1)-Z^2\alpha_S^2}{r^2}\Psi + 2Z\frac{\alpha_S}{r}\Psi, \tag{128}$$

$$-\widetilde{H}_{4r}\Psi = \frac{\partial^2 \Psi}{\partial R^2} - \frac{R^2}{R_0^4}\Psi - \frac{1}{R^2}\left(\frac{\partial^2 \Psi}{\partial \alpha^2} - \frac{1}{4}\Psi + \frac{l_s(l_s+1)}{\cosh^2 \alpha}\Psi\right), \quad l_s \le l = 0, 1, 2, \ldots, \quad (129)$$

$$-\widetilde{H}_{4i}\Psi = -i\frac{Z\alpha_S}{r^2}2\left(\frac{\partial \Psi}{\partial \alpha} - \tanh \alpha\, \Psi\right) - \frac{R^2 Z^2 \alpha_S^2}{r^4}\Psi. \quad (130)$$

We shall prove that the modified 4D energy $\widetilde{E}_\tau^{(4)}$ of the hydrogen atom is that of stationary oscillating states. The term $\widetilde{H}_{4r}$ accounts for the internal space-time rotations of the electron. In the absence of an external electromagnetic field, the electron is in the ground state $l_s = j_s - 1/2 = 0$. However, in the presence of the electromagnetic field the electron can be in an excited state $l_s = j_s - \frac{1}{2} = 0, \ldots, l$ induced by the spin-orbit orbit coupling. Hence, each momentum level $l_s$ splits up the Bohr levels into $l+1$ sublevels of order $\alpha_S^4$ representing the fine structure. The first term of the Hamiltonian (134) accounts for the spin-orbit coupling, which in our approach is the coupling of the time-like component of the intrinsic four-dimensional angular momentum of the electron with the external electric field. The scalar coupling of the squares, represented by the second term, dominates at short distances of the electron from the proton.

In the stationary case, we can rewrite the 4D Hamiltonian in the form

$$-\widetilde{H}_4 \Psi = \frac{\partial^2 \Psi}{\partial r^2} + \frac{2}{r}\frac{\partial \Psi}{\partial r} + k^2 \Psi - \frac{l(l+1) - Z^2 \alpha_S^2 + iZ\alpha_S h_{l_s}}{r^2}\Psi +$$

$$2Z\frac{\alpha_S}{r}\Psi + \frac{\partial^2 \Psi}{\partial R^2} - \frac{R^2}{R_e^4}\Psi = i\widetilde{E}_\tau^{(4)}\Psi, \qquad \frac{1}{R_e^2} = \sqrt{\frac{1}{R_0^4} + \frac{Z^2 \alpha_S^2}{r^4}}. \quad (131)$$

$$h_{l_s} = 2\int_{-\infty}^{\infty} \psi_{l_s}^\dagger(\alpha)\frac{\partial \psi_{l_s}(\alpha)}{\partial \alpha}\cosh \alpha\, d\alpha = (-1)^{l_s}(2l_s+1), \quad l_s \le l = 0, 1, \ldots. \quad (132)$$

We obtain the expectation value $h_{l_s}$ for each intrinsic angular state $l_s$ by substituting the relation (105) for $\psi_{l_s}(\alpha)$ in the integrand of (132). For the 2s-state $l_0 = 0$ we obtain $h_0 = 1$ and for the 2p-state $l_1 = 1$ we obtain $h_1 = -3$. The imaginary term $iZ\alpha_S h_{l_s}$ in (131) accounts for the slight jiggling called "*Zitterbewegung*" of the bound electron [17]. We decouple (131) approximately by the ansatz $\Psi(r, R) = \psi_r(r)\Psi_R(R)$, giving the decoupled equations

$$\frac{d^2\psi_r}{dr^2} + \frac{2}{r}\frac{d\psi_r}{r} + k^2\psi_r + 2Zk\frac{\alpha_S}{r}\psi_r + \frac{Z^2\alpha_S^2 - l(l+1) + iZ\alpha_S h_{l_s}}{r^2}\psi_r - k_e^2\psi_r = i\widetilde{E}_\tau^{(4)}\psi_r, \quad (133)$$

$$k_e^2 = \sqrt{k_C^4 + \frac{Z^2\alpha_S^2}{r^4}} = k_C^2 + k_r^2, \qquad k_r^2 = k_r^2(r) = \frac{Z^2\alpha_S^2}{k_C^2 r^2}\frac{1}{r^2 + \sqrt{r^4 + Z^2\alpha_S^2 k_C^{-4}}}, \quad (134)$$

$$\frac{\partial^2 \Psi_R}{k_e^2 \partial R^2} + (1 - k_e^2 R^2)\Psi_R = 0. \tag{135}$$

Relation (134) shows that the radius $R_e = 1/k_e$ of the 4D moment of inertia of the electron depends on the local electric field strength. As consequence, the effective mass $k_e \hbar/c$ known as the *renormalized* mass depends on the electric field strength. The origin of the effective mass of the electron is the same as that in a solid. The larger the electric field strength is the smaller is the 4D radius $R_e$ and the larger is the mass. This behavior may be the reason for the failure to determine the radius of the electron by scattering experiments. Within an atom, the radius is the smallest for the ground state. The term $k_r^2$ accounts for the change of the square of the mass within the electric field.

For solving (133), it is advantageous to write $\psi_r$ in the form

$$\psi_r = r^\sigma e^{-\kappa r} f(r). \tag{136}$$

Substituting (136) for $\psi_r$ in (133) we obtain

$$f'' + 2\left(\frac{a}{r^2} + \frac{\sigma+1}{r} - \kappa\right)f' + + \left(\kappa^2 - k_C^2 + k^2 - i\tilde{E}_\tau^{(4)}\right)f -$$
$$2\frac{\kappa(\sigma+1) - Z\alpha_S k}{r}f - \frac{Z^2\alpha_S^2 - l(l+1) + iZ\alpha_S h l_S - \sigma(\sigma+1)}{r^2}f - k_r^2(r)f = 0. \tag{137}$$

In order that $\psi_r$ satisfies the asymptotic condition $\lim_{r \to \infty} \psi_r = 0$, the function $f$ must not increase exponentially at the limit $r \to \infty$. This is only the case for a set of eigenvalues of the free exponentials $k, \tilde{E}_\tau^{(4)}, \kappa, \sigma$. Unfortunately, we cannot readily determine these eigenvalues because we cannot solve equation (137) analytically. However, we can determine the eigenvalues approximately with a sufficient degree of accuracy by replacing in (137) $k_r^2(r)$ by its expectation value

$$\langle k_r^2 \rangle = \int_0^\infty \psi_r^* k_r^2(r) \psi_r r^2 dr \approx \frac{Z^2\alpha_S^2}{k_C^2} \int_0^\infty \frac{\psi_{nl}^2(r)}{r^2 + \sqrt{r^4 + Z^2\alpha_S^2 k_C^{-4}}} dr. \tag{138}$$

We have obtained the final approximation by substituting the Schrödinger radial wave function $\psi_{nl}(r)$ for $\psi_r$ in the integrand of the first integral, $n = n_r + l + 1$ is the principal quantum number. The integral for the states $l > 0$ is negligibly small compared with that for the $s$-states ($l = 0$). For these states primarily the region $0 \leq r \leq \varrho_0 = \alpha_S^{1/2} Z^{1/2}/k_C$ contributes to the integral, giving

$$\int_0^\infty \frac{\psi_{nl}^2(r)dr}{r^2 + \sqrt{r^4 + \varrho_0^4}} \approx \frac{\psi_{n0}^2(0)}{\varrho_0} \delta_{0l} \int_0^\infty \frac{dx}{x^2 + \sqrt{x^4 + 1}} \approx \frac{5\delta_{0l}}{4\varrho_0} \psi_{n0}^2(0), \tag{139}$$

$$\langle k_r^2 \rangle \approx \frac{5 Z^{\frac{3}{2}} \alpha_S^{\frac{3}{2}}}{4 k_C} \psi_{n0}^2(0) \delta_{nl} = \frac{5(n-1)!}{n^2 [n!]^3} Z^{\frac{9}{2}} \alpha_S^{\frac{9}{2}} k_C^2 \delta_{0l}, \quad \delta_{0l} = \begin{cases} 1 & l = 0 \\ 0 & l > 0 \end{cases}. \tag{140}$$

By substituting (140) for $k_r^2(r)$ in (137), we obtain

$$f'' + 2\left(\frac{a}{r^2} + \frac{\sigma+1}{r} - \kappa\right) f' + \left(\kappa^2 - \langle k_e^2 \rangle + k^2 - i\tilde{E}_\tau^{(4)}\right) f -$$

$$2\frac{\kappa(\sigma+1) - Z\alpha_S k}{r} f + \frac{\sigma(\sigma+1) - L}{r^2} f = 0, \tag{141}$$

$$L = l(l+1) - Z^2 \alpha_S^2 - iZ\alpha_S h_{ls}, \quad \langle k_e^2(n,l) \rangle = k_C^2 + \frac{5(n-1)!}{n^2 [n!]^3} Z^{\frac{9}{2}} \alpha_S^{\frac{9}{2}} k_C^2 \delta_{0l}. \tag{142}$$

The solution of (141) is a polynomial for a set of eigenvalues of the parameters $k, \kappa, \sigma, \tilde{E}_\tau^{(4)}$. The exponents $k$ and $k_\tau$ are real whereas the exponents $\kappa = \kappa_r + i\kappa_r$ and $\sigma = \sigma_r + i\sigma_i$ are complex to account for the complex coefficient of the third and fifth term in (141). As a result, $f$ is a complex polynomial. In order to find this polynomial we first choose $\kappa_r$ and $\kappa_i$ such that the third term in (141) nullifies, which is the case if

$$\kappa_r^2 - \kappa_i^2 = k_e^2 - k^2, \quad 2\kappa_r \kappa_i = \tilde{E}_\tau^{(4)}. \tag{143}$$

In the next step, we determine the free exponent $\sigma = \sigma_r + i\sigma_i$ by postulating that the last term in (141) vanishes, giving the complex condition

$$L - \sigma(\sigma+1) = 0. \tag{144}$$

The solution for the real part $\sigma_r$ and for the imaginary part $\sigma_i$ of the exponent $\sigma$ are

$$\sigma_r = \sqrt{\tfrac{1}{2}\left(a + \sqrt{a^2 + b^2}\right)} - \tfrac{1}{2}, \quad \sigma_i = \tfrac{b}{\sqrt{2}}\left(a + \sqrt{a^2 + b^2}\right)^{-1/2}, \tag{145}$$

$$a = \left(l + \tfrac{1}{2}\right)^2 - Z^2 \alpha_S^2, \quad b = Z\alpha_S h_{l_s}. \tag{146}$$

For the $s$-states ($l = 0$, $l_s = 0$, $h_0 = -1$), we obtain

$$\sigma_{0r} = 0, \qquad \sigma_{0i} = Z\alpha_S. \tag{147}$$

By taking into account the relations (143) and (144) and introducing the dimensionless complex variable $2r\kappa = x$, equation (133) reduces to

$$xf'' + (2(\sigma + 1) - x)f' + vf = 0, \tag{148}$$

$$v = \frac{Z\alpha_s k}{\kappa} - \sigma - 1, \quad x = 2r\kappa. \tag{149}$$

The solution of (148) is a polynomial if the coefficient $v = v_r + iv_i$ is a positive integer number $v = n_r = 0, 1, 2, \ldots$ . It follows from (149) that this is the case if the imaginary part $v_i$ and the real part $v_r$ fulfill the relations

$$v_i = -\sigma_i - Z\alpha_s k \frac{\kappa_i}{\kappa_r^2 + \kappa_i^2} = 0, \tag{150}$$

$$v_r = -\sigma_r - 1 + Z\alpha_s k \frac{\kappa_r}{\kappa_r^2 + \kappa_i^2} = n_r. \tag{151}$$

The solution of (148) for integer $n_r$ is the generalized Laguerre polynomial $L_{n_r}^{(2\sigma+1)}(x)$ of order $n_r$ and index $2\sigma + 1$:

$$f(n_r, \sigma, r\kappa) = L_{n_r}^{(2\sigma+1)}(2r\kappa) = \sum_{\mu=0}^{n_r}(-2)^m \binom{n_r + 2\sigma + 1}{n_r - \mu}\frac{(r\kappa)^\mu}{\mu!}, \quad n_r = 0, 1, \ldots . \tag{152}$$

Because the parameters $\sigma_l$ and $\kappa$ are complex, the polynomial (152) is composed of two real polynomials, one representing the real part the other the imaginary part. This result resembles that obtained by the Dirac formalism, which also yields two radial polynomials for each energy state of the hydrogen atom [12, 17]. However, these polynomials differ from the real part and the imaginary part of the complex polynomial (152). Moreover, the radial density $\psi_r \psi_r^*$ obtained by our method stays finite at the origin for all states $n_r$ whereas the Dirac formalism yields for the *s*-state $n_r = 0$ a diverging density at the origin.

By employing the relations (150), (151) and (143), we eventually obtain the eigenvalues of $k, \kappa_r, \kappa_i, \tilde{E}_\tau^{(4)}$ as functions of the quantum numbers $l, l_s$ and $n_r$:

$$\kappa_r(n_r, l, l_s) = Z\alpha_S k \frac{\sigma_r + 1 + n_r}{\sigma_i^2 + (\sigma_r + 1 + n_r)^2}, \quad \kappa_i(n_r, l, l_s) = Z\alpha_S k \frac{\sigma_i}{\sigma_i^2 + (\sigma_r + 1 + n_r)^2}, \quad (153)$$

$$\tilde{E}_\tau^{(4)}(n_r, l, l_s) = Z^2 \alpha_S^2 k_C^2 \frac{2\sigma_i(\sigma_r + 1 + n_r)}{\left(\sigma_i^2 + (\sigma_r + 1 + n_r)^2\right)^2 + Z^2 \alpha_S^2 [(\sigma_r + 1 + n_r)^2 - \sigma_i^2]}, \quad (154)$$

$$E/\hbar c = k(n_r, l, l_s) \approx \sqrt{\frac{\langle k_e^2(n,l) \rangle}{1 + Z^2 \alpha_S^2 \frac{(\sigma_r + 1 + n_r)^2 - \sigma_i^2}{\left(\sigma_i^2 + (\sigma_r + 1 + n_r)^2\right)^2}}}. \quad (155)$$

The energy eigenvalues $E(n_r, l, l_s) = \hbar c k(n_r, l, l_s)$ of the hydrogen atom depend on the radial quantum number $n_r$, the orbital quantum number $l$, and the spin precession quantum number $l_s = j_s - 1/2$ of the electron, whereas the eigenvalues derived by the Dirac formalism depend on the two quantum numbers $n'$, $j = l \pm \frac{1}{2}$. According to our approach, the 4D energy $\tilde{E}_\tau^{(4)}(n_r, l, l_s)$ accounts for the creation of mass due to the bonding of the electron. Moreover, the eigenvalues (155) are not degenerated such that the energies for states with principal quantum number $n = n_r + l + 1$ are identical, as it is the case for the energies obtained by the Schrödinger equation. The origin of this difference is that within the frame of our theory the time-like intrinsic rotation of the electron interacts with the time-like (electric) component of electromagnetic field tensor whereas the space-like spin of the electron only interacts with the magnetic field.

### 13. Lamb shift

According to the Dirac theory, the energy of 2*s*-state of the hydrogen atom coincides with the energy of the 2*p*-state. However, contrary to the Dirac theory, our result shows that the 2*s*-state ($n_r = 2, l = l_s = 0$) and the 2*p*-state ($n_r = l = l_s = 1$) are not degenerated in agreement with the experiments of Lamb and Retherford [18]. Their result shows a slight separation of these energy states, the 2s-state being the higher of the two. Accordingly, the radiative transition is from the 2*s*-state to the 2*p*-state. Several eminent theoretical physicists have tried to explain the Lamb shift. They considered the Lamb shift as a radiative correction and employed the Feynman formalism of quantum electrodynamics (QED) for their calculations. Contrary to this procedure, our method attributes the change of the mass within the electromagnetic field as the origin of the Lamb shift.

For the $p$-states ($l = 1, l_s = 1, h_{11} = -3$), the real part and the imaginary part of the exponent $\sigma$ are

$$\sigma_{1r} = 1, \quad \sigma_{1i} = Z\alpha_S. \tag{156}$$

By employing these values to the electron energy (155) for the 2s-state ($n_r = 1, l = l_s = 0$) and for the 2p-state ($n_r = 0, l = 1, l_s = 1$) of the hydrogen atom ($Z = 1$) and considering that $\langle k_e^2(2, 1) \rangle = 0$, we obtain

$$\frac{E_{s2}}{\hbar c} = \frac{\sqrt{\langle k_e^2(2,0) \rangle}}{\sqrt{1+\alpha_S^2 \frac{4-\alpha_S^2}{(4+\alpha_S^2)^2}}}, \quad \frac{E_{p2}}{\hbar c} = \frac{k_C}{\sqrt{1+\alpha_S^2 \frac{4-\alpha_S^2}{(4+\alpha_S^2)^2}}}. \tag{157}$$

Note that (157) does not hold for the state $l = 1, l_s = 0$. If we neglect the renormalization of the mass of the bound electron by substituting $k_C^2$ for $\langle k_e^2(2, 0) \rangle$ in (161) we find that the 2s-state and the 2p state ($l = 1, l_s = 1$) are degenerated sharing the same energy as predicted by the Dirac theory. However, this prediction is in conflict with the experimental result obtained by Lamb and Retherford [18], which shows that the states are not degenerated. On the other hand, according to our result (157) the energies of the two states differ if we take the "bonding term" $\langle k_r^2 \rangle = \langle k_e^2 \rangle - k_C^2$ into account, giving

$$\Delta E_{\mathrm{L}} = E_{s2} - E_{p2} \approx \frac{5}{64} \alpha_S^{9/2} mc^2. \tag{158}$$

According to our result (158), the Lamb shift is proportional to $\alpha_S^{9/2}$ whereas that obtained by QED is proportional to $\alpha_S^5$. Nevertheless, both approaches give about the same value for $\Delta E_{\mathrm{L}}$. We have obtained the Lamb shift by introducing the canonical angular momentum (118). The Dirac theory only considers the bilinear term resulting from of the square of the canonical angular momentum operator accounting for the tensor interaction of the 4D angular momentum of the electron with the external electric field. Because the missing scalar term of the interaction yields the Lamb shift, the Dirac formalism cannot provide it. We have obtained the approximate result (158) for the Lamb shift, the Dirac by employing a straightforward procedure without needing the cumbersome procedure of quantum-field theory. In order to derive a more precise result, we must solve equation (137) numerically. Nevertheless, our approximate result agrees with the prediction of QED and the experimental results showing that the 2s-state is the higher of the two states. Moreover, our procedure gives a precise insight in the bonding structure of the electron within the hydrogen atom.

## 14. The proton problem

Experiments in the late 1960s at SLAC and CERN have shown that the charge density of the proton is not uniform but concentrated in discrete regions of its volume. The Standard model explains this result by postulating that every baryon is composed of three (massive) quarks with fractional charge. Unfortunately, hadrons absolutely lock up the quarks because one has never detected them in any experiment. We assume that nature uses a binary method for the construction of elementary particles although our assumption is in conflict with the Standard model. In order to be in agreement with the experimental results and to show the physical relevance of our hypothesis, we consider a massive particle formed by a massless quark (quarkino) and its antiparticle having opposite 4D helicity. The mutual 4D density of the quark-antiquark pair produces the potential (81), which represents a hyper-symmetrical quantum well with infinitely high walls. The potential attractive at distances $R > R_q$ and repulsive at distances $R < R_q$ enabling a large number of mass eigenstates.

By assuming that the massive compound particle has time-like spin $\mu = 1/2$, its radial wave equation (41)

$$\frac{d^2\psi_R}{dR^2} + \frac{3}{R}\frac{d\psi_R}{dR} - \left\{\frac{R}{R_q^3} + \frac{1}{RR_q} - \Lambda\right\}\psi_R + \frac{3}{4R^2}\psi_R = 0. \tag{159}$$

We reduce (159) by introducing the normalized radius $x = R/R_q$ and the modified wave function $\Psi_R(x) = R^{3/2}\psi_R(x)$, giving

$$\frac{d^2\Psi_R}{dx^2} - \left\{x + \frac{1}{x} - \Lambda R_q^2\right\}\Psi_R = 0. \tag{160}$$

In order that the curvature $\Psi_R''$ does not diverge at the origin, the wave function $\Psi_R(x)$ must nullify. In addition, the wave function must vanish at infinity to enable normalization. The equation (160) has for a set of eigenvalues $\Lambda_n$ solutions, which fulfill the boundary conditions

$$\Psi_R(\infty) = 0, \quad \Psi_R(0) = 0. \tag{161}$$

Unfortunately, we cannot determine these eigenvalues analytically because we can only solve the equation (160) numerically. However, we obtain approximate accurate eigenvalues because we can determine analytically the eigenfunctions in the region $x > 1$ for aarbitrary values. By imposing the constraint $\Psi_R(0) = 0$ on the asymptotic solutions, we obtain the approximate eigenfunctions in the entire range $0 \leq x \leq \infty$. Within the region $x \gg 1$, we can neglect the term $1/x$ in (160). The solution, which satisfies the condition $\Psi_R(\infty) = 0$, is the Airy function

$$\Psi_R(x) = \text{Ai}\,(x - \Lambda R_q^2), \tag{162}$$

The Airy function $\text{Ai}(z)$ is of great importance for numerous diffraction problems. It oscillates for negative arguments and decreases exponentially for positive arguments $z > 0$. We determine the eigenvalues $\Lambda = \Lambda_n$ by imposing on the asymptotic solution (16) the additional constraint $\Psi_R(0) = 0$, giving the eigenvalue equation

$$\text{Ai}\,(-\Lambda R_q^2) = 0. \tag{163}$$

The zeros $-a_n$ of the Airy function [9] determine the eigenvalues

$$\Lambda_n = \frac{a_n}{R_q^2}, \quad n = 1, 2, \dots\,. \tag{164}$$

Although the approximate eigensolutions $\Psi_{Rn}(x) = \text{Ai}\,(x - \Lambda_n R_q^2)$ satisfy the boundary conditions, their curvature differs from that of the exact solution in the region $0 \leq x \leq 1$. In particular, the curvature of the exact wave function at the origin $\Psi_R''(0) = \Psi_R'(0)$ is nonzero whereas that of the approximation is zero.

The lowest eigenvalue $\Lambda_1 \approx 2.34/R_q^2$ relates to the stable ground state of the particle. The maximum of the corresponding eigenfunction is located at radius $R_m \approx 1.3 R_q$, which is larger than the radius $R_{\min} = R_q$ of the minimum of the potential (81). The reason for this difference is that the curvature of the approximate eigenfunction nullifies at the origin $R = 0$, whereas that of the true eigenfunction is nonzero and positive shifting the maximum of the eigenfunction towards the potential minimum. We can determine the radius $R_q$ from the rest mass $m_p$ of the particle by applying the Compton relation $k_C = \frac{mc}{\hbar} = \sqrt{\Lambda_1} \approx 1.52/R_q$. The lowest eigenstate represents the ground that of the stable massive particle. Because the electron and the proton are the only stable massive elementary particles, we assign the radial wave function (163) to the proton. The 4D radial density of the excited proton consists of hyper-spherical shells whose number increases with increasing excitation or eigenvalue $\Lambda_n$, respectively. Experiments of deep inelastic scattering with electrons and protons show that the density of the exited proton is not uniform but has an internal structure (quarks). Because these findings agree with the predictions of our model, we conclude that the proton is very likely composed of two massless spin-1/2 quarks with opposite 4D helicity having non-fractional charges. The eigenstates are mass-eigenstates because the eigenvalues (164) are proportional to the square of the mass. Accordingly, any transition from an excited state to a lower state results in the emission of two massive particles with opposite 4D helicity, which subsequently decay in their stable constituents. Although our model differs from that of the Standard model, it has the advantage that the formalism is consistent starting from the supposition that massive elementary particle consist of massless sub-particles and their antiparticles.

## 15. Conclusions

Within the frame of our model elementary particles behave like four-dimensional spinning tops propagating in the four-dimensional Minkowski space. We represent the intrinsic 4D rotation of the particle by an antisymmetric tensor, which differs from the axial vector describing the orbital motion of the particle in the three-dimensional space. We define the 4D angular momentum of the massive particle by three angular eigenvalues and by the eigenvalue of its moment of inertia, which is the square of its mass. Our model suggests that photons and basic massless quarks (quakinos) are the basic elements of all particles. In particular, electrons and positrons are each composed of two photons. Feynman suggested in his space-time approach that we can treat a positron propagating forward is equivalent to an electron that propagates "backward in time" [19]. Our findings differ from this interpretation by showing that the sign of the space-time rotation $\alpha$ determines the sign of the electric charge.

Experiments show that most elementary particles have the tendency to decay into lighter particles, apart from the proton, the electron, and their antiparticles. However, even the electron and the positron can annihilate each other giving photons. The expectation value $\langle \vec{\Omega}_t \rangle$ of the time-like component (31) of the 4D angular tensor operator $\overleftrightarrow{\Omega}$ is not zero. For spin-1/2 particles, such as the electron and the positron, the time-like component $\vec{\Omega}_t$ of the 4D angular momentum has the property of an electric monopole as documented by the expectation value

$$\langle \vec{\Omega}_t \rangle = \pm \frac{1}{2} \vec{e}_r. \tag{175}$$

We assign the positive sign to the positron and the negative sign to the electron. Hence, the expectation value of the time-like component of the 4D angular momentum accounts for the sign of the charge of an electric monopole in the three-dimensional space. Therefore, these particles do not possess an electric dipole moment, as proven by the precise measurements by J. Baron et al. [20]. Their finding is in agreement with our theoretical result, which shows that the imaginary space-time rotation is the origin of the charge. The expectation value $\langle \vec{\Omega}_s \rangle = s\vec{e}_z, s = \pm 1/2$ of the space-like component of the angular tensor operator defines the orientation of the spin in the 3D space. In the presence of an external magnetic field, the direction $\vec{e}_z$ of the spin is either parallel or anti-parallel to the magnetic field strength $\vec{B} = B\vec{e}_z$. Our relativistic approach shows that the time-like and the space-like components of the 4D angular momentum of electrons and positrons are the source of their charge (electric monopole) and their magnetic moment (spin), respectively. The approach elucidates the nature of the wave-particle dualism of elementary particles by describing each particle wave as a 4D vortex superposed on the carrier wave like a surfer riding on the surf. The interaction of the vortex with the detector produces the recorded signal, not the carrier wave as postulated in classical quantum mechanics. Because our approach yields the Lamb shift in a straightforward way and elucidates its origin, we believe that the novel theory is realistic. Moreover, our model provides the attractive 4D potential for quark confinement in a relativistic consistent way, whereas the Standard model employs a heuristic three-dimensional potential.

Our theoretical investigations elucidate the formation of massless elementary particles and of stable massive elementary particles composed of a massless particle and its antiparticle. Their opposite 4D helicity induces an attraction forming a massive compound particle such as the electron or the proton. In order to investigate the structure of compound particles consisting of more than two massive subunits, such as the neutron, we must solve the three-body problem in the 4D space. In principle, the novel theory allows one to tackle this formidable problem because we can extend the theory to systems consisting of many particles. Unfortunately, we do not know anything about the associated 4D potentials and forces and the role of neutrinos, which are constituents of almost all massive particles. Nevertheless, our approach can describe the decay of particles, at least in principle, because the time-dependent fundamental equation (4) accounts for the creation and annihilation of mass, as documented in detail for the neutrino oscillations provided they exist.

# Appendix

## I. Four-dimensional vector and tensor analysis

### 1. Introduction

The Minkowski space is a four-dimensional pseudo-Euclidian space put up by the four orthogonal coordinates $x_1 = x, x_2 = y, x_3 = z, x_4 = ict$. In the three-dimensional case, the unit vectors $\vec{e}_1 = \vec{e}_x$, $\vec{e}_2 = \vec{e}_y$, $\vec{e}_3 = \vec{e}_z$ define the directions of the coordinates. These unit vectors satisfy the relations

$$\vec{e}_\mu \cdot \vec{e}_\nu = \delta_{\mu\nu}, \quad \vec{e}_\mu \times \vec{e}_\nu = -\vec{e}_\nu \times \vec{e}_\mu, \quad \delta_{\mu\nu} = \begin{cases} 1 & \mu = \nu \\ 0 & \mu \neq \nu \end{cases}. \tag{1}$$

The vector product $\vec{e}_\mu \times \vec{e}_\nu$ of two unit vectors is an axial unit vector, which is orthogonal with respect to the plane put up by the two vectors. By going from three to four dimensions, we define four unit vectors $\vec{e}_\mu, \mu = 1, 2, 3, 4$, each of which defines one of the four orthogonal directions in Minkowski space. In the three-dimensional space, the vector product of two different unit vectors is an axial unit vector, which points in the direction of the third unit vector. This behavior differs from that of the unit vectors in the four-dimensional space because 12 two-dimensional surfaces enclose the four-dimensional cube resulting in six different directions of their normal vectors. In order to define the direction of these normal vectors, we must replace the three-dimensional vector products by four-dimensional dyadic products $\vec{e}_\mu \vec{e}_\nu$, which satisfy the relations

$$\vec{e}_\mu \vec{e}_\nu + \vec{e}_\nu \vec{e}_\mu = 2\delta_{\mu\nu}, \quad \mu, \nu = 1, 2, 3, 4. \tag{2}$$

In accordance with the three-dimensional formalism, we interpret the dyadic product $\vec{e}_\mu \vec{e}_\nu$ for $\mu = \nu$ as a scalar product and as a vector product in the case $\mu \neq \nu$. Our approach is in accordance with the Clifford algebra and gives it a geometrical interpretation. The dyadic products satisfy the same commutation rule as the Dirac matrices $\gamma_\mu, \mu = 1, 2, 3, 4$. However, within the frame of the Dirac theory each $\gamma_\mu$ represents a 4x4 matrix.

The four-dimensional cube in Minkowski space consists of four three-dimensional cubes, three of which are of space-time nature. The pseudo dyadic vectors $\vec{e}_1\vec{e}_2\vec{e}_3$, $\vec{e}_1\vec{e}_2\vec{e}_4$, $\vec{e}_2\vec{e}_3\vec{e}_4$, $\vec{e}_3\vec{e}_1\vec{e}_4$ determine the location of the 3D cubes in Minkowski space. By employing the 4D unit vectors, we write the four-dimensional vector potential with components $A_\mu$ as

$$\vec{A} = \sum_{\mu=1}^{4} \vec{e}_\mu A_\mu. \tag{3}$$

The anti-symmetric four-dimensional second-rank tensor $\overleftrightarrow{T}$ has six components $T_{\mu\nu} = -T_{\nu\mu}$. This four-dimensional tensor is the equivalent of an axial vector in the three-dimensional space. Using the vector notation, we obtain the representation

$$\overleftrightarrow{T} = \sum_{\mu,\nu} \vec{e}_\mu \vec{e}_\nu T_{\mu\nu} = 2 \sum_{\mu=1}^{4} \sum_{\nu>\mu}^{4} \vec{e}_\mu \vec{e}_\nu T_{\mu\nu}. \tag{4}$$

## 2. Dyadic four-dimensional gradient acting on vectors and tensors

In order to derive the electromagnetic field tensor from the four-dimensional vector potential we must extent the concepts of three-dimensional vector analysis to four dimensions. The most important differential operator is the 4-gradient

$$\text{Grad} = \sum_{\mu=1}^{4} \vec{e}_\mu \frac{\partial}{\partial x_\mu}. \tag{5}$$

Applying the four-dimensional gradient (5) to a scalar function is rather trivial. We obtain a more involved result by performing the gradient of a four-dimensional vector $\vec{A}$, giving

$$\text{Grad}\,\vec{A} = \sum_{\mu,\nu} \vec{e}_\mu \vec{e}_\nu \frac{\partial A_\nu}{\partial x_\mu} = \sum_{\mu=1}^{4} \frac{\partial A_\mu}{\partial x_\mu} + \sum_{\mu=1}^{4} \sum_{\nu>\mu}^{4} \vec{e}_\mu \vec{e}_\nu \left( \frac{\partial A_\nu}{\partial x_\mu} - \frac{\partial A_\mu}{\partial x_\nu} \right). \tag{6}$$

The first term in the last relation is a scalar quantity defining the 4D divergence

$$\text{Div}\,\vec{A} = \sum_{\mu=1}^{4} \frac{\partial A_\mu}{\partial x_\mu}. \tag{7}$$

The last term in (6) is the electromagnetic field tensor $\overleftrightarrow{F}$ with components

$$F_{\mu\nu} = -F_{\nu\mu} = \frac{\partial A_\nu}{\partial x_\mu} - \frac{\partial A_\mu}{\partial x_\nu}. \tag{8}$$

Because the anti-symmetric tensor $\overleftrightarrow{F}$ describes a rotation in 4D space, it represents the 4D curl of the 4D vector potential

$$\text{Curl}\,\vec{A} = \frac{1}{2}\overleftrightarrow{F} = \frac{1}{2}\sum_{\mu,\nu=1}^{4} \vec{e}_\mu \vec{e}_\nu F_{\mu\nu} = \sum_{\mu=1}^{4} \sum_{\nu>\mu}^{4} \vec{e}_\mu \vec{e}_\nu F_{\mu\nu}. \tag{9}$$

By introducing the definitions (7) and (9), we rewrite the 4D-gradient (6) in the form

$$\text{Grad}\vec{A} = \text{Div}\vec{A} + \text{Curl}\vec{A}. \tag{10}$$

The expression (10) reveals that the four-dimensional gradient operation of al 4-vector yields a scalar and an anti-symmetric second-rank tensor.

In the next step we determine the 4-D gradient of (10), giving

$$\text{Grad}\,\text{Grad}\vec{A} = \text{Grad}\,\text{Div}\vec{A} + \text{Grad}\,\text{Curl}\vec{A}. \tag{11}$$

In order to elucidate the structure of the second term on the right-hand side of (11), we perform the differentiations, giving

$$\text{Grad}\,\text{Curl}\vec{A} = \frac{1}{2}\text{Grad}\overleftrightarrow{F} = \frac{1}{2}\sum_{\lambda=1}^{4} \vec{e}_\lambda \sum_{\mu,\nu=1}^{4} \vec{e}_\mu \vec{e}_\nu \frac{\partial F_{\mu\nu}}{\partial x_\lambda} =$$

$$-\sum_{\mu,\nu=1}^{4} \vec{e}_\mu \frac{\partial F_{\mu\nu}}{\partial x_\nu} + e_0 \sum_{\mu=1}^{4} \vec{e}_\mu (-)^\mu \sum_p \frac{\partial F_{\nu\kappa}}{\partial x_\lambda}, \quad e_0 = \vec{e}_1 \vec{e}_2 \vec{e}_3 \vec{e}_4. \tag{12}$$

The pseudo scalar $e_0$ is the dyadic product of the four unit vectors whereas the equivalent quantity $\gamma_0$ of the Dirac formalism represents a 4x4 matrix. The summation index $p$ in the last sum denotes the permutation of the indices $\nu \neq \mu, \kappa \neq \mu, \nu \neq \mu$, where $\nu, \kappa, \lambda$ are any three of the integers 1, 2, 3, 4. The sum of each of the four permutations for fixed $\mu = 1, 2, 3, 4$ is

$$\sum_p \frac{\partial F_{\nu\kappa}}{\partial x_\lambda} = \frac{\partial F_{\nu\kappa}}{\partial x_\lambda} + \frac{\partial F_{\lambda\nu}}{\partial x_\kappa} + \frac{\partial F_{\kappa\lambda}}{\partial x_\nu}. \tag{13}$$

The structure of the relation (12) shows that the 4-gradient of an anti-symmetric second-rank tensor ($F_{\mu\nu} = -F_{\nu\mu}$) is composed of a 4-vector and a pseudo 4-vector, which represents an entirely anti-symmetric tensor of the third rank.

### 3. Derivation of Maxwell's equations

As an example for the physical relevance of the 4D vector analysis we choose for $\vec{A}$ the four-dimensional electromagnetic vector potential with components

$$A_1 = A_x, \quad A_2 = A_y, \quad A_3 = A_z, \quad A_4 = \frac{i}{c}\varphi. \tag{14}$$

The space-like components $A_\mu$, $\mu = 1, 2, 3$ are those of the three-dimensional magnetic vector potential. The imaginary time-like component $A_4$ represents the scalar electric potential $\varphi$. The 4D electromagnetic vector potential satisfies the Lorentz gauge

$$\text{Div}\vec{A} = \sum_{\mu=1}^{4} \frac{\partial A_\mu}{\partial x_\mu} = 0. \tag{15}$$

The components of the electromagnetic field tensor

$$F_{12} = \frac{\partial A_2}{\partial x_1} - \frac{\partial A_1}{\partial x_2} = B_x, \quad F_{13} = \frac{\partial A_3}{\partial x_1} - \frac{\partial A_1}{\partial x_3} = -B_y, \quad F_{23} = \frac{\partial A_3}{\partial x_2} - \frac{\partial A_2}{\partial x_3} = B_z,$$

$$F_{14} = \frac{\partial A_4}{\partial x_1} - \frac{\partial A_1}{\partial x_4} = \frac{i}{c}E_x, \quad F_{24} = \frac{\partial A_4}{\partial x_2} - \frac{\partial A_2}{\partial x_4} = \frac{i}{c}E_y, \quad F_{34} = \frac{\partial A_4}{\partial x_3} - \frac{\partial A_3}{\partial x_4} = \frac{i}{c}E_z. \tag{16}$$

We obtain the Maxwell equations from (12) and (15)) by postulating that

$$\sum_p \frac{\partial F_{\nu\kappa}}{\partial x_\lambda} = \sum_p \frac{\partial}{\partial x_\lambda}\left(\frac{\partial A_\kappa}{\partial x_\nu} - \frac{\partial A_\nu}{\partial x_\kappa}\right) = 0, \tag{17}$$

$$\sum_{\mu,\nu=1}^{4} \vec{e}_\mu \frac{\partial F_{\mu\nu}}{\partial x_\nu} = \sum_{\mu,\nu=1}^{4} \vec{e}_\mu \frac{\partial}{\partial x_\nu}\left(\frac{\partial A_\nu}{\partial x_\mu} - \frac{\partial A_\mu}{\partial x_\nu}\right) = -\sum_{\mu,\nu=1}^{4} \vec{e}_\mu \frac{\partial^2 A_\mu}{\partial x_\nu^2} = \mu_0 \vec{J}, \tag{18}$$

$$\vec{J} = \sum_{\mu=1}^{4} \vec{e}_\mu J_\mu = \vec{j} + \vec{e}_4 j_4, \quad \vec{j} = \rho_e \frac{d\vec{r}}{dt} = \rho_e \vec{v}, \quad j_4 = i\rho_e c. \tag{19}$$

We have obtained the last relation in (18) by considering the Lorentz gauge (15); $\rho_e$ is the three-dimensional charge density and $\vec{\beta} = \vec{v}/c$. The absolute value of the 4D current density $\vec{J}$ is invariant under Lorentz transformations. By considering the relations (16), we rewrite (17) and (18) in the form

$$\text{curl}\vec{B} - \frac{1}{c^2}\frac{\partial \vec{E}}{\partial t} = \mu_0 \vec{j}, \quad \text{div}\vec{E} = \mu_0 c \rho_e = \rho_e/\varepsilon_0 , \tag{20}$$

$$\text{curl}\vec{E} + \frac{\partial \vec{B}}{\partial t} = 0, \quad \text{div}\vec{B} = 0. \tag{21}$$

$$\sum_{\nu=1}^{4} \frac{\partial^2 A_\mu}{\partial x_\nu^2} = -\mu_0 J_\mu, \quad \mu = 1, 2, 3, 4. \tag{22}$$

The vector $\vec{j} = \rho_e \vec{v}$ is the three-dimensional current density. Equation (22) forms a set of four equations, which are equivalent in all respects to the standard Maxwell equations (20) and (21). By taking into account the relations (15) and (17), we can replace the set of Maxwell equations by the single equation

$$\text{Grad}(\text{Grad}\,\vec{A}) = \mu_0 \vec{J}. \tag{23}$$

Moreover, we can simplify the resulting equations by imposing on the vector potential the Lorentz gauge

$$\text{Div}\vec{A} = 0. \tag{24}$$

## II. Space-time rotation and Lorentz transformations

The Lorentz transformation represents a rotation of the four-dimensional coordinate system in one of the three two-dimensional space-time planes of the Minkowski space. Each of these planes embeds the time-like coordinate $x_4$ and one of the spatial coordinates $x_\nu, \nu = 1, 2, 3$. The rotation by the angle $\chi$ transforms the two coordinates $x_4 = ict, x_\nu$ into the coordinates $\tilde{x}_4, \tilde{x}_\nu$ of the rotated coordinate system. Without loss of generality, we assume that the two systems of coordinates coincide at zero time and that their common origin is located at position $x_4 = \tilde{x}_4 = 0, x_\nu = \tilde{x}_\nu = 0$. In this case, we readily obtain the relations between the coordinates of the two systems by writing the rotational transformation in complex form as

$$\tilde{x}_4 + i\,\tilde{x}_\nu = (x_4 + i\,x_\nu)e^{i\chi}. \tag{1}$$

The separation of (1) in two equations, one containing the real parts the other the imaginary parts, gives

$$\tilde{x}_4 = x_4 \cos\chi - x_\nu \sin\chi = \frac{1}{\sqrt{1+\tan^2\chi}}(x_4 - x_\nu \tan\chi), \tag{2}$$

$$\tilde{x}_\nu = x_4 \sin\chi + x_\nu \cos\chi = \frac{1}{\sqrt{1+\tan^2\chi}}(x_\nu + x_4 \tan\chi). \tag{3}$$

The $\tilde{x}_4, \tilde{x}_\nu$ system moves uniformly with velocity $v = v_\nu = dx_\nu/dt$ relative to the $x_4, x_\nu$ system in the direction parallel to the $x_\nu$ axis. We have shown in the article (15) that the uniform velocity $v = v_\nu$ and the imaginary angel $\chi = i\alpha$ of the space-time rotation satisfy the relation

$$\tan\chi = i\tanh\alpha = i\beta = i\frac{v}{c}, \quad -\infty \leq \alpha \leq \infty. \tag{4}$$

By substituting $i\beta = 1v/c$ for $\tan\chi$ in (2) and (3) and considering the relations $\tilde{x}_4 = ic\tilde{t}$, $x_4 = ict$, we readily obtain the Lorentz transformations

$$\tilde{t} = \frac{1}{\sqrt{1-\beta^2}}\left(t - x_\nu \frac{\beta}{c}\right), \quad \tilde{x}_\nu = \frac{1}{\sqrt{1-\beta^2}}(x_\nu - c\beta t). \tag{5}$$

The interaction of the time-like angular moment (intrinsic electric field) of the electron with the electric field of the proton introduces a Lorentz-invariant 4D oscillation with frequency $\omega_\tau = D_0 \tilde{E}_\tau^{(4)}$, regardless of the inertial system. The result convincingly proves that the introduction of the universal time is realistic because our approach shows that the electron is encircling the positive nucleus with a velocity, which depends on its orbit in accordance with classical mechanics and the Bohr model. The orbital velocity and the universal time $\tau$ are Lorentz-invariant quantities, which are not accessible to the Schrödinger and Dirac formalisms because these consider only states, which do not depend on the universal time $\tau$. Because these theories cannot describe dynamical processes in the 4D space, we consider them as incomplete.